\documentclass[twocolumn]{aastex62}

\usepackage{subfigure}

\graphicspath{{./}{figures/}}

\revised{\today}
\submitjournal{ApJ}

\usepackage{graphicx}	
\usepackage{amsmath}	
\usepackage{amssymb}	
\usepackage{natbib}
\usepackage{hyperref}

\newcommand{\LISA}{\textit{LISA}}

\newcommand{\LIGOVirgo}{LIGO/Virgo}

\newcommand{\Gaia}{\textit{Gaia}}
\newcommand{\cosmic}{\textsc{COSMIC}}

\begin{document}


\title[COSMIC Variance in Binary Population Synthesis]{COSMIC Variance in Binary Population Synthesis}

\correspondingauthor{Katelyn Breivik}
\email{kbreivik@cita.utoronto.ca}

\author[0000-0002-9660-9085]{Katelyn Breivik}
\affil{Canadian Institute for Theoretical Astrophysics, University
of Toronto, 60 St. George Street, Toronto, Ontario, M5S 1A7,
Canada}
\author[0000-0002-0403-4211]{Scott Coughlin}
\affil{Center for Interdisciplinary Exploration and Research in Astrophysics (CIERA), Northwestern University, 
\\1800 Sherman Road, Evanston, IL 60201, USA}
\affil{Department of Physics and Astronomy, Northwestern University, 2145 Sheridan Road, Evanston, IL 60208, USA}
\affil{Physics and Astronomy, Cardiff University, Cardiff, CF10 2FH, UK}
\author[0000-0002-0147-0835]{Michael Zevin}
\affil{Center for Interdisciplinary Exploration and Research in Astrophysics (CIERA), Northwestern University, 
\\1800 Sherman Road, Evanston, IL 60201, USA}
\affil{Department of Physics and Astronomy, Northwestern University, 2145 Sheridan Road, Evanston, IL 60208, USA}
\author[0000-0003-4175-8881]{Carl L. Rodriguez}
\affil{Harvard Institute for Theory and Computation, 60 Garden St, Cambridge, MA 02138, USA}
\author[0000-0002-4086-3180]{Kyle Kremer}
\affil{Center for Interdisciplinary Exploration and Research in Astrophysics (CIERA), Northwestern University, 
\\1800 Sherman Road, Evanston, IL 60201, USA}
\affil{Department of Physics and Astronomy, Northwestern University, 2145 Sheridan Road, Evanston, IL 60208, USA}
\author[0000-0001-9582-881X]{Claire S. Ye}
\affil{Center for Interdisciplinary Exploration and Research in Astrophysics (CIERA), Northwestern University, 
\\1800 Sherman Road, Evanston, IL 60201, USA}
\affil{Department of Physics and Astronomy, Northwestern University, 2145 Sheridan Road, Evanston, IL 60208, USA}
\author[0000-0001-5261-3923]{Jeff J. Andrews}
\affil{Center for Interdisciplinary Exploration and Research in Astrophysics (CIERA), Northwestern University, 
\\1800 Sherman Road, Evanston, IL 60201, USA}
\affil{Department of Physics and Astronomy, Northwestern University, 2145 Sheridan Road, Evanston, IL 60208, USA}
\affil{Niels Bohr Institute, University of Copenhagen, Blegdamsvej 17, 2100 Copenhagen, Denmark}
\author{Michael Kurkowski}
\affil{Department of Physics, University of Notre Dame, 225 Nieuwland Science Hall, Notre Dame, IN 46556, USA}
\affil{Center for Interdisciplinary Exploration and Research in Astrophysics (CIERA), Northwestern University,
\\1800 Sherman Road, Evanston, IL 60201, USA}
\author[0000-0003-3815-7065]{Matthew C. Digman}
\affil{Center for Cosmology and AstroParticle Physics (CCAPP), Ohio State University, Columbus, OH 43210, USA}
\author{Shane L. Larson}
\affil{Center for Interdisciplinary Exploration and Research in Astrophysics (CIERA), Northwestern University,
\\1800 Sherman Road, Evanston, IL 60201, USA}
\affil{Department of Physics and Astronomy, Northwestern University, 2145 Sheridan Road, Evanston, IL 60208, USA}
\author[0000-0002-7132-418X]{Frederic A. Rasio}
\affil{Center for Interdisciplinary Exploration and Research in Astrophysics (CIERA), Northwestern University,
\\1800 Sherman Road, Evanston, IL 60201, USA}
\affil{Department of Physics and Astronomy, Northwestern University, 2145 Sheridan Road, Evanston, IL 60208, USA}

\begin{abstract}
The formation and evolution of binary stars is a critical component of several fields in astronomy. The most numerous sources for gravitational wave observatories are inspiraling and/or merging compact binaries, while binary stars are present in nearly every electromagnetic survey regardless of the target population. Simulations of large binary populations serve to both predict and inform observations of electromagnetic and gravitational wave sources. Binary population synthesis is a tool that balances physical modeling with simulation speed to produce large binary populations on timescales of days. We present a community-developed binary population synthesis suite: COSMIC which is designed to simulate compact-object binary populations and their progenitors. As a proof of concept, we simulate the Galactic population of compact binaries and their gravitational wave signal observable by the Laser Interferometer Space Antenna (\LISA). 
\end{abstract}



\section{Introduction}\label{sec:Intro}
Binary stars play a critical role, either as a signal or noise source, in nearly all fields of astronomy \citep{whitepaper}. Binary systems containing stellar remnants are the most prolific sources for both ground- and space-based gravitational wave (GW) observatories. The Laser Interferometer Gravitational Wave Observatory and Virgo (\LIGOVirgo) have detected the inspiral and merger of ten binary black holes (BHs) and one binary neutron star (NS) \citep[BNS;][]{GWTC-1}. The Laser Interferometer Space Antenna (\LISA) is expected to observe the population of ${\sim}10^7$ binary stellar remnants, or compact binaries, in the Milky Way and its surrounding environment forming a confusion foreground of gravitational radiation in the millihertz region of the GW spectrum.  Tens of thousands of compact binaries, dominated in number by double white dwarfs (DWDs), are expected to be resolved above the foreground, offering a unique probe of the populations of stellar remnants in the local Universe \citep[e.g. ][]{Nelemans2001a, Ruiter2010, Yu2010, Nissanke2012, Littenberg2013, Yu2015, Korol2017, Lamberts2018, Lamberts2019}. 

Simulations of compact-object binary populations are useful tools that enable astrophysical interpretations of GW sources and their progenitors. Binary population synthesis (BPS) combines single star evolution with prescriptions for binary interactions to simulate binary populations from zero age main sequence (ZMAS) through to the stellar remnant phase. Generally, each BPS study seeks to determine which physical processes are most important in shaping observed catalog sources. To this end, in each study a single population and its detection catalog are simulated for a single model or a set of several models that canvass the available parameter space.

\begin{figure*}
\label{fig: cosmic_schem}
\includegraphics[width=0.95\textwidth]{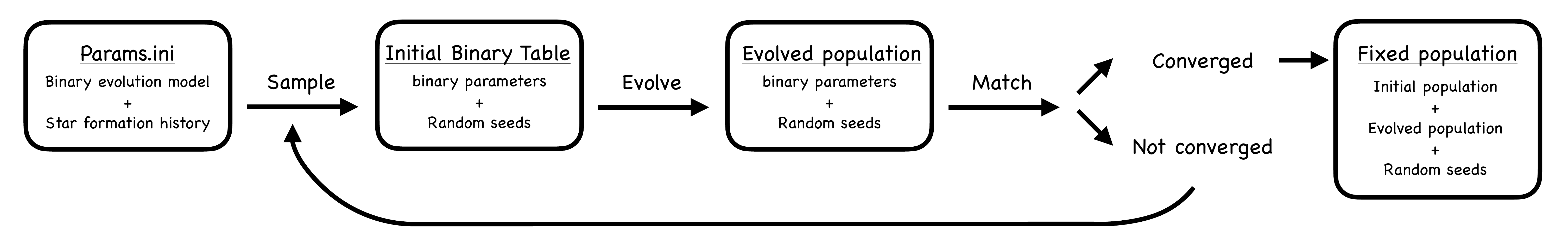}
\caption{Schematic for the process \cosmic\ uses to generate a fixed population. Generally, the process moves from left to right. All quantities in the boxes are produced and available to the user, while all arrows represent modules within \cosmic\ that facilitate the evolution process. For a discussion of the $match$ and convergence, see \autoref{sec: convergence}. See also the list of steps outlined below in Section\,\ref{sec:cosmic}.}
\end{figure*}

Several BPS codes are currently in use, each with their own simulation techniques and focuses. This type of simulation originated with the work of \cite{Whyte1985} and was soon followed by several BPS studies \citep[][e.g.]{Dewey1987, Lipunov1987, deKool1990, Hils1990, Ritter1991}. BPS methods have since been iterated upon and widely applied to study many different binary populations of interest. One approach is to modify a stellar evolution code previously developed for single stars to include the effects of binary evolution, as found in ev/STARS/TWIN \citep{Pols1995, Nelson2001}, the Brussels population number synthesis code, PNS \citep{DeDonder2004}, BINSTAR \citep{Siess2013}, MESA \citep{Paxton2011, Paxton2013, Paxton2015, Paxton2018, Paxton2019}, or BPASS \citep{Eldridge2016, Stanway2016, Eldridge2017, Stanway2018}. Another approach is to generate large libraries of lookup tables for single stars and then use interpolation combined with simple binary evolution prescriptions, as done in SEVN \citep{Spera2015, Spera2017, Spera2019} or \textsc{ComBinE} \citep{Kruckow2018}.

The most widely applied technique is to use fitting formulae derived from single star evolution models, which vary as a function of stellar age, mass, and metallicity. For example \cite{Hurley2000} developed such a formalism based on the stellar evolution models of \cite{Pols1998}. Different binary evolution prescriptions are applied to these fitting formulae to develop BPS codes, such as Scenario Machine \citep{Lipunov1996a, Lipunov1996b, Lipunov2009}, IBiS \citep[and references therein]{Tutukov1996}, \textsc{BSE} \citep{Hurley2002}, SeBa \citep{Portegies1996, Nelemans2001a, Toonen2012}, StarTrack \citep{Belczynski2002, Belczynski2008}, binary\_c \citep{Izzard2004,Izzard2006,Izzard2009}, COMPAS \citep{Stevenson2017, Barrett2018}, and MOBSE \citep{Giacobbo2018, Giacobbo2018b}. Given the wide variety of available software, studies like the \textsc{PopCORN} project, which compared the outputs of SeBa, StarTrack, binary\_c, and the Brussels PNS code, are an invaluable resource which quantify theoretical uncertainties and confirm results across BPS software \citep{Toonen2014}. More recent population synthesis tools which use Markov-Chain Monte Carlo methods \citep[{\tt dart\_board};][]{Andrews2018}, Gaussian processes \citep{Barrett2017, Taylor2018}, or Adaptive Importance Sampling \citep[STROOPWAFEL;][]{Broekgaarden2019} are also available to provide better statistical descriptions of binary populations.

Here we present \cosmic\ (Compact Object Synthesis and Monte Carlo Investigation Code), a BPS code adapted from BSE that generates large binary populations which can be convolved with star formation history (SFH) and spatial distribution models to produce astrophysical realizations of binary populations.

Simulated detection catalogs from these astrophysical realizations can then inform the range of possible compact-object binary populations detectable with both GW or electromagnetic observations. The ability to generate a statistical sample of binary populations is especially important when considering populations with low numbers. For example, the subset of mass transferring DWDs with helium accretors observable by \LISA\ \textit{and} \Gaia\ is expected to contain \textit{dozens} of systems, which are a tiny subset of the tens of thousands of DWDs individually detectable by \LISA\ \citep{Kremer2017,Breivik2018}. 

The paper is organized as follows. In Section~\ref{sec:cosmic} we summarize the key features of \cosmic. We detail the additional binary evolution prescriptions included in \cosmic\ beyond those contained in the original version of \textsc{BSE} in Section~\ref{sec:upgrades}. In Section~\ref{sec:fiducialMW} we demonstrate the capabilities of \cosmic\ to produce a reference population of compact binaries observable by \LISA\ and finish with conclusions in Section~\ref{sec:discussion}.  

\section{Overview: \cosmic}\label{sec:cosmic}
\cosmic\ is a python-based, community-developed BPS software suite with extensive documentation\footnote{https://cosmic-popsynth.github.io/}. \cosmic's binary evolution is based on \textsc{BSE}, but has been extensively modified to include updated prescriptions for massive star evolution and binary interactions (see \autoref{sec:upgrades}). All methods needed to generate a population, from initialization to scaling to astrophysical populations, are included in \cosmic. 

One of the main features of \cosmic\ is its ability to {\it adaptively\/} determine the size of a simulated binary population such that it adequately describes the population's parameter distributions based on the user's need. However, we note that \cosmic\ is also able to simulate a population with a predetermined size. All data used to generate a population of binaries, including the parameters of stochastic processes like natal kicks for compact objects, as well as the properties of the population itself, are saved in the output of \cosmic. This allows a user to analyze the entire population, as well as individual interesting systems, from ZAMS all the way to compact object formation. In the subsections below, we outline the process to simulate an astrophysical binary population using \cosmic.

\subsection{Fixed population}\label{sec: FixedPop}
The main output from \cosmic\ is the fixed population, a collection of binary systems that contains enough binaries to capture the underlying shape of the population's parameter distribution functions resulting from a user-specified star SFH and binary evolution model. The fixed population can be convolved with more complex SFHs and scaled to a large number of astrophysical populations. These populations can then be synthetically ``observed'' and used to explore the variance in synthetic catalogs associated with the simulated binary population and binary evolution model.

The general process to simulate a fixed population is as follows:
\begin{enumerate}
\item Select a binary evolution model and SFH.
\item Generate an initial population based on user-selected models for the SFH and initial binary parameter distributions.
\item Evolve the initial population according to the user-specified binary evolution model.
\item If on the first iteration, compare a subset containing half of the simulated population to the total population to determine how closely the binary parameter distributions match one another (we introduce a quantitative `match' condition for making this assessment, described in Sec\,\ref{sec: FixedPop}). If on the second or later iteration, compare the population from the previous iteration to the population containing both the current and previous iteration. 
\item If the binary parameter distributions have converged, the population is called the `fixed population' which contains a large enough sample to describe the essential statistical features of a binary evolution model.
\item Scale the fixed population to astrophysical populations, weighted either by mass or by number, by sampling the fixed population with replacement.
\item Apply a synthetic observation pipeline to generate a set of synthetic catalogs from the astrophysical populations.
\end{enumerate}

Figure\,\ref{fig: cosmic_schem} illustrates the structure and process \cosmic\ uses to generate the fixed population. The fixed population is simulated once for each binary evolution model; thus a study with, for example, ten binary evolution models will have ten associated fixed populations. Astrophysical populations can then be sampled from the fixed population, convolved with a SFH, and used to generate a statistical set of synthetic catalogs for each model (i.e. a full study with ten binary evolution models and one SFH may contain \textit{ten thousand} synthetic catalogs).

\subsubsection{Initializing a population}\label{sec : Initialize}
The demonstration of \cosmic\ presented in this paper evolves binary populations using an adapted version of \textsc{BSE}, with several modifications including updated common envelope, wind, and kick prescriptions which are described further in Section~\ref{sec:upgrades}. Regardless of binary evolution model, the fixed population is generated from an initial population of binaries sampled from distribution functions to assign each binary with an initial metallicity ($Z$), primary mass ($m$), mass ratio ($q$), orbital separation ($a$), eccentricity ($e$), and birth time ($T_0$) according to a given SFH. Since binary evolution codes generally evolve a single binary at a time, initial binary populations can be generated using several different star formation histories and distribution functions. 

\cosmic\ is equipped to generate initial populations according to several binary parameter distributions. If the parameters are treated independently, initial masses may be sampled from a \citet{Salpeter1955}, \citet{Kroupa1993}, or \cite{Kroupa2001} initial mass function (IMF); mass ratios are uniformly sampled \citep{Mazeh1992,Goldberg1994}; orbital separations are sampled log-uniformly, according to {\"O}pik's Law, with lower limits set such that the ZAMS stars do not fill thier Roche lobes and an upper limit of $10^5\,\rm{R}_{\odot}$ following \citet{Dominik2012}; eccentricities may be sampled from a thermal distribution \citep{Heggie1975} or uniform distribution \citep{Geller2019}; binarity can be assumed to have user-specified fractions or the mass-dependent fraction of \citet{vanHaaften2013}. \cosmic\ is also able to generate initial binary samples following the \citet{Moe2017} multi-dimensional binary parameter distributions, which include mass and separation dependent binary fractions. For easy comparison to previous studies, we use independently distributed parameters with primary masses following the \cite{Kroupa1993} IMF, a thermal eccentricity distribution and a constant binary fraction of $0.5$. For a study of the impact of multi-dimensional initial distributions on compact object populations see \cite{DeMink2015} and \cite{Klencki2018}.

\cosmic\ can generate binary populations according to very simple SFH prescriptions. For a more detailed study of the importance of SFH, see \cite{Lamberts2018}. In many cases, it is useful to initially choose a single burst of star formation and later convolve the population with a SFH appropriate for the astrophysical population of interest.

\subsubsection{Convergence of parameter distributions}\label{sec: convergence}
There is no formulaic way to \textit{a priori} predict the required number of binaries to be evolved for a fixed population, since each population depends on a different binary evolution model. The ideal number of simulated systems in a fixed population is such that the population adequately describes the final parameter distribution functions while not simulating so many systems as to be inefficient. To quantify this number, we develop a discrete \textit{match} criteria, based on the cross ambiguity function, or overlap function, used in matched filtering techniques \citep[e.g. Eq. 6 of][]{Chatziioannou2017}.

We use independently generated histograms for each binary parameter, with binwidths determined using Knuth's Rule \citep{Knuth2006} implemented in \textsc{astropy} \citep{Astropy2013}, to track the distribution of each parameter as successive populations are simulated and cumulatively added to the fixed population. Before each histogram is generated, we ensure that similar binwidths are used for each parameter by transforming each set of binary parameters to lie between 0 and 1. We then enforce the physical limits of the simulated systems (e.g. positive definite values for mass and orbital period and eccentricities between $0$ and $1$) by taking the inverse logistic transform, or logit, which redistributes the transformed data to have limits between -$\inf$ and $\inf$.

We define the $match$ as
\begin{equation}
\label{eq:match}
match = \frac{\sum\limits_{k=1}^{N} P_{k,i} P_{k,i+1}}{ \sqrt{ \sum\limits_{k=1}^{N} (P_{k,i}P_{k,i})\sum\limits_{k=1}^{N} (P_{k,i+1}P_{k,i+1})}},
\end{equation}
\noindent where $P_{k,i}$ denotes the probability for the $k$th bin for the $i$th iteration. The $match$ is limited to values between 0 and 1 and tends to unity as the parameter distributions converge to a distinct shape. The $match$ is specified by the user, and can be set to any value. For the study in Section~\ref{sec:fiducialMW}, we set $match \geq 1-10^{-5}$ as a fiducial choice, but caution that the results of each simulated population should be carefully checked to confirm that the population does not contain artificial gaps due to low-number statistics in cases of very rare sub-populations.

\begin{figure*}
    \centering
    \includegraphics[width=0.95\textwidth]{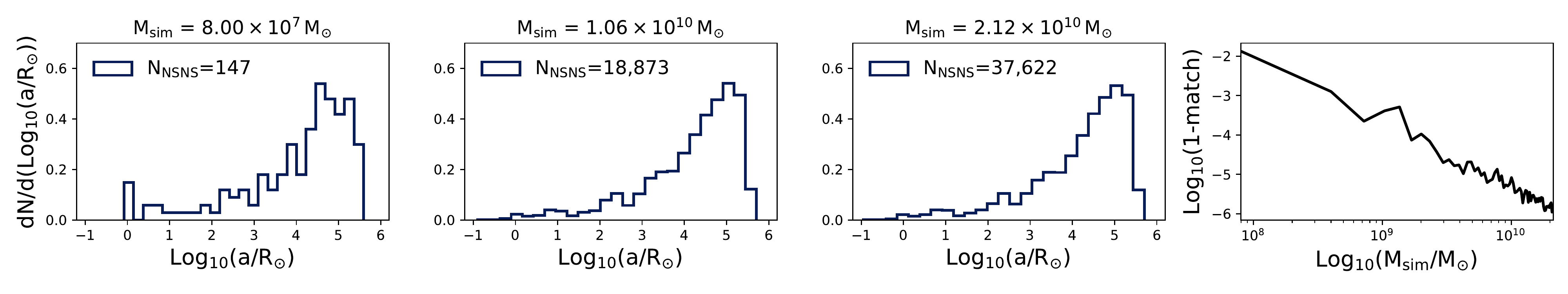}
    \caption{The first three columns show normalized histograms of the semimajor axis at formation for simulated NS+NS populations where each column includes the population from the previous ones. The fifth column shows the evolution of the $match$ as the size of the simulated population grows, where we show Log$_{10}$(1-$match$) to illustrate how the $match$ tends to unity. } 
    \label{fig:match}
\end{figure*}

Figure\,\ref{fig:match} uses the solar metallicity NS + NS population from \autoref{sec:fiducialMW} below to illustrate how distributions of the semimajor axis converge to a distinct shape as more binaries are simulated. The evolution of the $match$ (see Equation\,\ref{eq:match}) as a function of total simulated mass is also shown in terms of Log$_{10}$(1-$match$) to quantify the convergence of the distributions. As the total simulated mass increases, more NS + NS systems are added to the population which fills in the normalized distribution and consequently drives the $match$ toward one.

\subsection{Generating astrophysical population realizations}\label{sec:MCsample}
Once the fixed population satisfies the user-specified convergence criteria, an astrophysical population can be sampled and each binary in the astrophysical population can be assigned a position and orientation. The number of sources in each astrophysical population can be calculated by normalizing the size of the fixed population by the ratio of the mass of the astrophysical population to the mass of stars formed to produce the fixed population:
\begin{equation}
    N_{\rm{astro}} = N_{\rm{fixed}}\frac{M_{\rm{astro, tot}}}{M_{\rm{fixed, stars}}},
\end{equation}
\noindent or by the ratio of the number of stars in the astrophysical population to the total number of stars formed to produce the fixed population:
\begin{equation}
    N_{\rm{astro}} = N_{\rm{fixed}}\frac{N_{\rm{astro, tot}}}{N_{\rm{fixed, stars}}}.
\end{equation}

For each astrophysical realization, $N_{\rm{astro}}$ binaries are sampled with replacement from the fixed population. These sampled binaries can also be assigned a three dimensional position and an inclination ($i$), an argument of periapsis ($\omega$), and a longitude of the ascending node ($\Omega$).  

\cosmic\ allows for several general Galactic position distributions. 
For the axisymmetric thin and thick disks, the radial and vertical distributions are assumed to be independent. We adopt Galactic position distributions and Galactic component masses from \citet{McMillan2011} for the thin disk, thick disk, and bulge as a fiducial model. However, \cosmic\ can distribute binaries according to exponential distributions in the radial and vertical directions with any scale height, as well as in spherically symmetric distributions. 

Following \citet{McMillan2011}, we assume the mass in the thin and thick disks to be $4.32\times10^{10}\,M_{\odot}$ and $1.44\times10^{10}\,M_{\odot}$ respectively, while we assume the mass of the bulge to be $8.9\times10^{9}\,M_{\odot}$. 
Finally, we emphasize that while \cosmic\ has utility functions implemented to distribute binaries in the Milky Way, the spatial distributions are independent of the binary evolution prescriptions. Thus multiple spatial distributions can be assumed for a single fixed population and more sophisticated spatial distribution models can be integrated into future releases of \cosmic. 

\section{Updates to BSE}\label{sec:upgrades}

\cosmic\ uses a modified version of BSE to evolve binaries from ZAMS through to compact object formation. We describe several upgrades to binary evolution prescriptions contained in \cosmic\ below. The version containing these upgrades is fixed as \textsc{v3.2.0} \citep{COSMIC_3.2}. All future modifications will be openly developed in the \cosmic\ Github repository.\footnote{https://github.com/COSMIC-PopSynth/COSMIC}

\subsection{Winds}\label{subsec:winds}

Mass loss through stellar winds plays an important role in compact object formation because this determines the mass of the star just before it undergoes a supernova (SN) explosion \citep[e.g.,][]{Mapelli2009,Belczynski2010,Fryer2012,Giacobbo2018}. In recent years, models of stellar winds have been revised to reflect updates in our understanding of various relevant physical processes. We have amended the original \textsc{BSE} stellar wind prescriptions with several of these prescriptions, summarized below.

Recent work has shown that line-driven winds exhibit strong dependence upon metallicity both on the main sequence (MS) and post-MS \citep[e.g.,][]{Vink2001,MeynetMaeder2005,GrafenerHamann2008,Vink2011}. We have updated the original \textsc{BSE} prescription to include metallicity-dependent winds for O and B stars as well as for Wolf-Rayet stars. The winds for O and B stars are treated according to the prescription of \citet{Vink2001}, which considers stars with temperatures $12,500\,\rm{K} < T_{\rm{eff}} < 22,500\,\rm{K}$ and $27,500\,\rm{K} < T_{\rm{eff}} < 50,000\,\rm{K}$ separately. As in \citet{Rodriguez2016b}, we adopt the methods of \citet{Dominik2013} where the high and low temperature prescriptions are extended to $T_{\rm{eff}}=25,000\,\rm{K}$. Wolf-Rayet star winds are treated according to \citet{Vink2005}.

Additionally, recent models suggest that as stars approach the limit imposed by the electron-scattering Eddington factor, $\Gamma_e = \chi_e L / (4 \pi c G M)$ where $\chi_e$ is electron scattering opacity, winds may become insensitive to metallicity \citep[e.g.,][]{GrafenerHamann2008,Vink2011,Chen2015}. Thus, we include Eddington-limited winds using the prescriptions described in \citet{GrafenerHamann2008,Giacobbo2018}.

If the primary star loses mass through a stellar wind, the secondary may accrete some of the ejected material as it orbits through it. The accretion rate onto the secondary can be estimated according to a Bondi-Hoyle type mechanism \citep{BondiHoyle1944}, which is sensitive to the velocity of the wind lost from the primary, $v_{\rm{W}}^2 = 2\beta_{\rm{W}} G M/R$, where $\beta_{\rm{W}}$ is a constant depending on stellar type \citep[see equation 9 of][]{Hurley2002}. We have implemented $\beta_{\rm{W}}$ values from \citet{Belczynski2008}, and choose this as a default. In this case, $\beta_{\rm{W}}$ is a function of stellar type. For H-rich MS stars, $\beta_{\rm{W}}=0.125$ for masses below $1.4\,M_{\odot}$, $\beta_{\rm{W}}=7.0$ for masses above $120\,M_{\odot}$, and is linearly interpolated between. For H-rich giants, $\beta_{\rm{W}}=0.125$ regardless of mass. For He-rich stars, including MS and giants, $\beta_{\rm{W}}=0.125$ for masses below $10\,M_{\odot}$, $\beta_{\rm{W}}=7.0$ for masses above $120\,M_{\odot}$, and is linearly interpolated between.

\subsection{Mass transfer stability and common envelope}\label{subsec:CE}
The stability of Roche-lobe overflow mass transfer is determined through the same process as in \cite{Hurley2002}, using critical mass ratios determined from radius-mass exponents \citep{Webbink1985} where $q_{\rm{crit}} = m_{\rm{donor}}/m_{\rm{accretor}}$. COSMIC allows both models for critical mass ratios for giant star donors detailed in Section~(2.6.1) of \cite{Hurley2002}. We have added new models that are consistent with binary\_c, following \cite{Claeys2014} where we have modified the critical mass ratios of MS/helium-MS stars and binaries with degenerate accretors based on \citet{deMink2007}. For the models which are consistent with \cite{Claeys2014}, we adopt the Roche-lobe overflow mass transfer rates in their Eq.~(10) in the case of stable mass transfer. We have also added a model that is consistent with StarTrack following \cite{Belczynski2008}. The most notable difference in the critical mass ratios between StarTrack and both BSE or binary\_c is the treatment of helium-MS donors where $q_{\rm{crit}} = 1.7$ and for helium stars on the Hertzsprung Gap or giant branch where $q_{\rm{crit}} = 3.5$ based on the findings of \cite{Ivanova2003}. Finally, we note that \cosmic\ allows for user-specified critical mass ratios based on evolutionary stages to allow easy integration with future models as they arise. As a fiducial model for critical mass ratios, we choose the standard model reported in \cite{Hurley2002} which does not differentiate between models for degenerate and non-degenerate accretors. 

We employ the standard $\alpha\lambda$ model for common envelope (CE) evolution as done in \cite{Hurley2002}. In this case, systems which undergo unstable mass transfer enter into a CE which can be expelled by the injection of orbital energy from the binary. In this formalism, $\lambda$ is a factor which determines the binding energy of the envelope to its stellar core, while $\alpha$ is the efficiency factor for injecting orbital energy into the envelope. As with the currently available version of \textsc{BSE}, \cosmic\ defaults to a variable $\lambda$ which depends on the evolutionary state of the star following the description in the Appendix of \citet{Claeys2014}. However, constant $\lambda$ are also allowed as an option. 

As a fiducial CE model, we use the variable binding energy parameter in conjunction with a constant CE efficiency parameter $\alpha=1.0$ following previous results \citep[e.g.,][]{Nelemans2001a,Dominik2012}. However, previous studies of post-CE binaries point to an efficiency as low as $\alpha=0.2$ \citep{Zorotovic2010,Toonen2013,Camacho2014}. 
On the other hand, detailed modeling of the CE phase for DNS progenitors suggests CE efficiencies may be as high as $\alpha \approx 5$ \citep{Fragos2019}, which may also reduce the tension between rate predictions from CE channels \citep[e.g.,][]{Mapelli2018} and the empirical DNS merger rate derived from \LIGOVirgo\ \citep{GW170817}. 

We also include an option for a ``pessimistic CE'' scenario, in which unstable mass transfer from a donor star without a well-developed core-envelope structure is always assumed to lead to a merger \citep{Belczynski2008}. This assumption applies to hydrogen rich and helium stars on the Main Sequence and Hertzsprung Gap, as well as all white dwarfs.
When this option is set, we also assume mergers to occur when unstable mass transfer is triggered from donor stars without a clear entropy jump at the core-envelope boundary \citep{Ivanova2004}, which includes stars on the Hertzsprung Gap. Note, however, that the outcomes of CEs in this case is still uncertain \citep[e.g.][]{Deloye2010}.

Mass transfer involving an evolved He-star donor and a compact object is a critical phase of binary evolution for forming hardened compact binaries that can merge within a Hubble time as GW sources. 
For progenitors of DNSs, detailed modeling of this phase finds that mass transfer typically proceeds stably \citep{Ivanova2003,Tauris2013,Tauris2015} and does not lead to the onset of a CE. 
Stable mass transfer is also corroborated by studies that compare population modeling to the properties of Galactic DNS \citep{Vigna-Gomez2018}. 
Post-mass-transfer separations are typically wider if this phase is modeled stably versus unstably. 
We allow for the stable mass-transfer evolution of He-star donors with compact object companions to be approximated using the fitting formulae in \cite{Tauris2015}, which can fit for either the post-mass-transfer separation and remaining He-star envelope mass, or just the post-mass-transfer separation. 

One important change implemented in \cosmic\ is the treatment of supernovae that directly follow a CE phase. 
By default, if a supernova immediately follows a CE (which typically occurs for evolved He-star donors), \textsc{BSE} determines the post-SN barycentric velocity, orbital properties, and survival using the \textit{post-CE} orbital separation and \textit{pre-CE} stellar mass. 
Therefore, the mass-loss in the supernova includes the mass of the donor-star envelope that formed the CE --- a self-inconsistent treatment since the ejection of this envelope is used to determine the hardening during the CE phase. 
This inconsistency leads to artificially amplified mass-loss kicks during the supernova, which is particularly apparent in extremely tight binaries. 
Therefore, by default, we use the post-CE separation and post-CE mass (which does not include the mass of the envelope) when determining the impact of the subsequent supernova. 
The impact of this change, particularly on DNS population properties, is explained in more detail in \cite{Zevin2019}. 

Finally, we note that contact systems can occur if both stellar components overflow their Roche radii or if a stellar component's radius is larger than the binary's periastron distance. In this case, a common envelope is always triggered, regardless of the mass transfer stability criteria. This simplification does not allow for long-lived contact binaries, e.g. W UMa, to be studied in detail, however their formation rates and progenitor populations can be investigated.

\subsection{Supernova explosion mechanisms and natal kicks}\label{sec:SNe}

\subsubsection{Standard Core-collapse Supernovae}\label{sec:CCSNe}

Two new prescriptions for the supernova mechanism have been added following \citet{Fryer2012}, which are both convection-enhanced and neutrino driven and account for material falling back onto the compact objects formed in core-collapse SNe. The two cases are delineated by the time between core bounce and explosion, with a `rapid' explosion which only allows for explosions that occur within $250\,\rm{ms}$ and a `delayed' explosion, which allows for longer timescales to explosion. The main difference in these two prescriptions is the presence of a mass gap between NSs and BHs which is produced in the rapid case but is not present in the delayed case. 

The inclusion of fallback onto compact objects reduces natal kick magnitudes due to the fraction of ejected mass during the supernova that falls back onto the compact object. 
As in the original version of \textsc{BSE}, natal kick magnitudes for standard iron core-collapse supernovae are drawn from a Maxwellian distribution with a dispersion of $265$\,km\,s$^{-1}$, consistent with the proper motions observed for isolated pulsars \citep{Hobbs2005}. 
The natal kick is then reduced by a factor of $1-f_{\rm fb}$, where $f_{\rm fb}$ is the fraction of the ejected supernova mass that will fall back onto the newly formed proto-compact object (see Eqs. (16) and (19) in \citealt{Fryer2012}). 
This efficiently damps the natal kicks for heavy ($M \gtrsim 30\,M_{\odot}$) BHs whose progenitors have CO core masses of $ M_{\rm CO} \geq 11\,M_{\odot}$. 
We also include options to allow for no natal kick, full natal kicks (with no reduction due to fallback), and ``proportional'' kicks that have BH kicks scaled down by a factor of $m_{\rm BH}/m_{\rm NS}$, where $m_{\rm NS}$ is the maximum mass of a NS (assumed to be $3.0\,M_{\odot}$). 
An option is also included to scale down the natal kicks for all BHs by a constant factor. This kick reduction is applied separately and simultaneously to any reduction due to fallback. We note that BHs and NSs are treated as `compact objects' in both \textsc{BSE} and \cosmic, where the difference between the two is chosen by the maximum NS mass.

By default, the natal kick is assumed to impact the proto-compact object in a direction that is isotropically sampled. 
However, correlations between the proper motions of pulsars and their spin axis suggest that the kick may be preferentially directed along the spin axis of a newly-formed compact object \citep{Wang2006, Ng2007, Kaplan2008}. 
We therefore allow for the kick direction to be constrained within a specified opening angle around the poles of the proto-compact object. 
Alternatively, the natal kick magnitudes and directions for a binary can be input directly when evolving a binary. 

\subsubsection{Electron-capture Supernovae}\label{subsubsec:ECSN}
A number of analyses have argued that for stars with main sequence masses in the range $\sim8-11 M_{\odot}$, the expected fate is a so-called electron-capture SN \citep[ECSN; e.g.,][]{Miyaji1980electron,Nomoto1984,Nomoto1987,Podsiadlowski2004,Ivanova2008}. In this scenario, stars develop helium cores in the range of masses $\sim1.5-2.5 M_{\odot}$ and never develop an iron core. In this case, collapse is triggered by electron captures onto $^{24}$Mg and $^{20}$Ne which lead to a sudden drop of electron pressure support in the stellar core. The collapse occurs when the mass of the stellar core is $>1.38\,\rm{M_{\odot}}$ \citep{Miyaji1980electron,Nomoto1984,Nomoto1987,Ivanova2008}.

The specific range of helium core masses expected to give rise to ECSNe is uncertain and several values have been proposed in the literature. In \citet{Hurley2002}, the range $1.6-2.25\,M_{\odot}$ was implemented. \citet{Podsiadlowski2004} argued that a broader range of $1.4-2.5\,M_{\odot}$ is more realistic. \citet{Belczynski2008} implemented the relatively narrow range of $1.85-2.25\,M_{\odot}$, while \citet{Andrews2015} argue that a range of $2-2.5\,M_{\odot}$ is required to reproduce the distribution of DNSs in the Milky Way field. Here we have updated the original \textsc{BSE} prescription to allow the lower and upper limits for the helium core mass range leading to ECSN to be specified directly. As a default, we follow \citet{Podsiadlowski2004}.

In the case of an ECSN, the SN is expected to occur through a prompt (fast) explosion rather than a delayed neutrino-driven explosion; thus various analyses have argued that ECSNe likely lead to smaller NS natal kicks relative to core-collapse SNe \citep[e.g.,][]{Podsiadlowski2004,Ivanova2008}. By default, we assume kicks resulting from ECSN are drawn from a Maxwellian with dispersion velocity $\sigma_{\rm{ECSN}}=20\rm{km/s}$, but include this as a variable to be specified directly by the user.

ECSNe may also occur through accretion-induced collapse (AIC) or merger-induced collapsed (MIC). Simple prescriptions are implemented for AIC and MIC in \textsc{BSE} \citep{Hurley2002}. For an ONeMg WD, if it is accreting CO or ONe material from its companion in a binary during RLOF, it will undergo AIC \citep{Nomoto1991condition,Saio2004off} when its mass is larger than the ECSN critical mass. Furthermore, a merger/collision between two CO or ONe WDs leads to a MIC if the mass of the merger/collision product is larger than the ECSN critical mass. Both paths can lead to NS remnants if the mass of the final product is smaller than the maximum NS mass set by \textsc{BSE}.

\subsubsection{Ultra-stripped Supernovae}\label{subsubsec:USSNe}
 
For close binaries containing a BH or NS and a Roche-lobe filling helium star companion, the helium star may be sufficiently stripped of material such that a naked $\sim1.5\,M_{\odot}$ core remains \citep{Tauris2013,Tauris2015}. The ensuing explosion of these ultra-stripped stars may lead to ejected mass $\lesssim 0.1\,M_{\odot}$, which may yield natal kick velocities far below those expected for standard core-collapse SNe. Thus, it may be appropriate to draw kicks from a Maxwellian with dispersion width, $\sigma$, smaller than that of the standard \citet{Hobbs2005} distribution ($\sigma=265$\,km\,s$^{-1}$). Here, whenever a helium star undergoes a CE phase with a compact companion such that a naked helium star forms, we implement the capability of assigning to these objects a smaller natal kick upon collapse and explosion as an ultra-stripped SNe.

\subsubsection{Pair-instability and Pulsational Pair-Instability Supernovae}\label{sec:PISN}

In the cores of post-carbon burning stars with sufficiently massive helium cores of $\gtrsim\,30\,M_{\odot}$, photons will readily convert into electron-positron pairs and diminish the pressure support of the core. 
This will cause the core to rapidly contract and the temperature to increase, allowing for the ignition of carbon, oxygen, or silicon \citep[e.g.][]{Woosley2015}. 
For helium-core masses of $\approx\,30-64\,M_{\odot}$, this spontaneous burning leads to mass ejections, known as pulsational pair instabilities (PPIs; \citealt{Woosley2017}). 
These will proceed until the instability is avoided. 
If the helium-core mass is in the range $\approx\,64-133\,M_{\odot}$, the instability exceeds the binding energy of the star and the star is completely destroyed --- a pair instability supernova (PISN; \citealt{Woosley2017}). 

As our default, we adopt the prescription from \cite{Belczynski2016}, which sets the maximum mass of the helium core below which the star is not destroyed by PISN to $45\,M_{\odot}$ (resulting in a remnant BH mass of $40.5\,M_{\odot}$, assuming 10\% of the mass is lost in the conversion from baryonic to gravitational mass). 
Helium core masses between $45-135\,M_{\odot}$ lead to the destruction of the star through a PISN, and therefore no remnant formation. 
We allow for the limiting helium-core mass beyond which a PISN occurs to be set manually to different values. 

Alternatively, multiple other prescriptions for determining the (P)PISN mass range and resultant remnant mass are available in \cosmic: 
\begin{enumerate}
    \item Prescription from \cite{Spera2017} (see Appendix~B), which is derived from fitting the masses of the compact remnants as a function of the final Helium mass fraction and final helium core mass from the simulations in \cite{Woosley2017}. 
    
    \item Fit to the grid of simulations from \cite{Marchant2019} (see Table 1), which demonstrate a turnover in the relation between pre-supernova helium core mass and final mass. 
    Similar to \cite{Stevenson2019}, we use a {$9^{\rm th}$-order} polynomial fit to map CO core masses between $31.99 \leq M_{\rm He}/\mathrm{M}_{\odot} \leq 61.10$ to BH masses: 
    \begin{equation}
        M_{\rm BH} = \sum_{l=0}^{8} c_l \left(\frac{M_{\rm He}}{M_{\odot}}\right)^l, 
    \end{equation}
    where the coefficients are
    $c_0=-6.29429263 \times 10^{5}$, 
    $c_1=1.15957797 \times 10^{5}$, 
    $c_2=-9.28332577 \times 10^{3}$, 
    $c_3=4.21856189 \times 10^{2}$, 
    $c_4=-1.19019565 \times 10^{1}$, 
    $c_5=2.13499267 \times 10^{-1}$, 
    $c_6=-2.37814255 \times 10^{-3}$, 
    $c_7=1.50408118 \times 10^{-5}$, and 
    $c_8=-4.13587235 \times 10^{-8}$.
    \footnote{We use different coefficients than in \cite{Stevenson2019} since we find that these are a better fit to data from \citet{Marchant2019} than those originally presented in \citet{Stevenson2019}.}
    The PISN gap, which leaves behind no remnant, is $54.48 < M_{\rm CO}/\mathrm{M}_{\odot} < 113.29$. 
    
    \item Fit to the grid of simulations from \cite{Woosley2019} (see Table 5), which agree reasonably well with those from \cite{Marchant2019} except that \cite{Woosley2019} find slightly lower helium core masses undergo PPISN. 
    We again fit a use a {$9^{\rm th}$-order} polynomial fit, mapping CO core masses between $29.53 \leq M_{\rm He}/\mathrm{M}_{\odot} \leq 60.12$ to BH masses, with coefficients
    $c_0=-3.14610870 \times 10^{5}$, 
    $c_1=6.13699616 \times 10^{4}$, 
    $c_2=-5.19249710 \times 10^{3}$, 
    $c_3=2.48914888 \times 10^{2}$, 
    $c_4=-7.39487537$, 
    $c_5=1.39439936 \times 10^{-1}$, 
    $c_6=-1.63012111 \times 10^{-3}$, 
    $c_7=1.08052344 \times 10^{-5}$, and 
    $c_8=-3.11019088 \times 10^{-8}$. 
   The PISN gap, which leaves behind no remnant,  is $60.12 < M_{\rm CO}/\mathrm{M}_{\odot} < 113.29$. 
    
\end{enumerate}

\subsection{Black Hole Spins}\label{subsec:spins}

We have added new prescriptions for the spins of newly-formed BHs from collapsing massive stars.  Due to the uncertain values associated with BH natal spins, we allow the spins of all BHs to be set to a specific Kerr value (specified by the user) or drawn from a uniform distribution whose bounds are also user specified.  In addition, we have included the prescriptions for BH spin based on the pre-collapse CO core mass of the progenitor from \cite{Belczynski2017}, which result in high spins for low-mass BHs ($\lesssim 30M_{\odot}$, depending on metallicity) and low spins for high-mass BHs.  

The latter prescription, based on stellar models computed by the \textsc{Geneva} stellar evolution code \citep{Eggenberger2008}, assumes that angular momentum is transported in massive stars via meridional currents, and is not efficient enough to spin down the core prior to collapse.  This is in contrast to newer work \cite[e.g.,][]{Fuller2019} suggesting that the Taylor-Spruit magnetic dynamo may allow for extremely efficient angular momentum transport through the envelopes of massive stars, producing BHs with dimensionless Kerr spin parameters $a/M \sim 0.01$. We do not include the latter prescription in COSMIC, nor do we allow for the BH spin to be increased by accretion or mergers.  These effects will be added in a future release.

\subsection{Pulsar Formation and Evolution}\label{subsec:pulsar}
We have updated \textsc{BSE} to implement the NS magnetic field and spin period evolution following \cite{Kiel2008pulsar} and \cite{ye2019msp}. All NSs are born with magnetic fields and spin periods that match the observed young pulsars \citep{Manchester2005catalogue}; their initial magnetic fields and spin periods are randomly drawn in the range of $10^{11.5}-10^{13.8}$ G, and $30-1000$ ms, respectively.

For single NSs or NSs in detached binaries, we assume magnetic dipole radiation for the spin period evolution. The spin-down rate is calculated by 

\begin{equation}
\Dot{P}=K\frac{B^2}{P},
\end{equation} 

\noindent where $P$ is the spin period, $B$ is the surface magnetic field and $K=9.87\times10^{-48}\,\rm{yr/G^2}$. We also assume that the magnetic fields follow exponential decays in a timescale of $\tau=3$ Gyr \citep{Kiel2008pulsar}, \begin{equation} 
B=B_0\exp\,\Big(\frac{-T}{\tau}\Big),
\end{equation} 

\noindent where $B_0$ is the initial magnetic field, and $T$ is the age of the NSs.

On the other hand, binary evolution will affect the NS magnetic fields and spin periods. For NSs in non-detached binaries, their magnetic fields can change significantly during mass accretion on short time scales. We assume ``magnetic field burying" \citep[e.g.,][]{Bhattacharya1991pulsar,Rappaport1995pulsar,Kiel2008pulsar,Tauris2012formation} for the magnetic field decay during mass accretion 

\begin{equation} 
B=\frac{B_0}{1+(\Delta M/10^{-6}M_{\odot})}\exp\,\Big(-\frac{T-t_{\rm{acc}}}{\tau}\Big).
\end{equation} 

\noindent $\Delta M$ is the mass accreted and $t_{\rm{acc}}$ is the accretion duration. These NSs are spun up according to the amount of angular momentum transferred \citep[][equation (54)]{Hurley2002}.

\begin{figure*}
    \centering
    \includegraphics[width=0.95\textwidth]{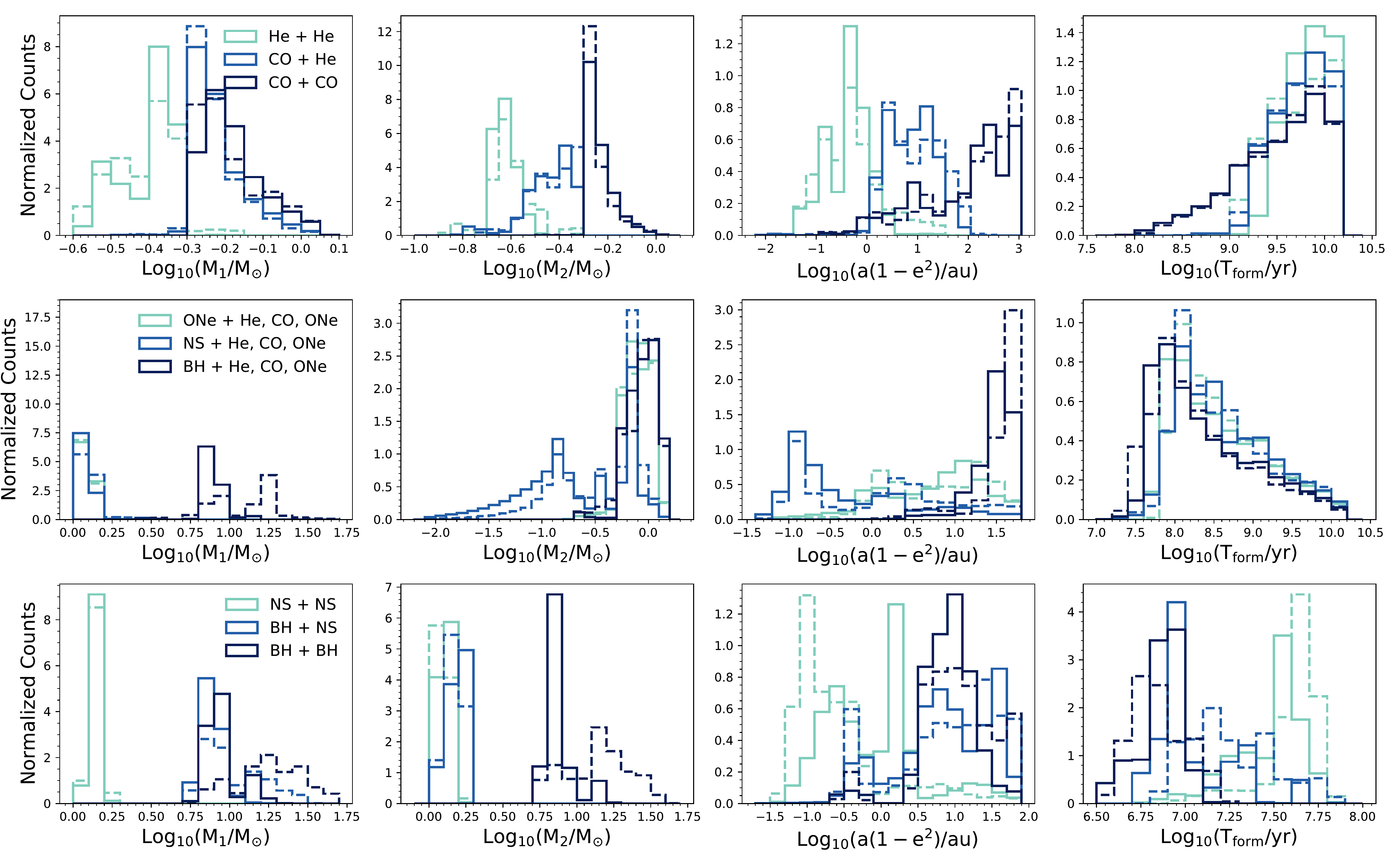}
    \caption{Normalized histograms of the primary mass, secondary mass, semilatus rectum, and formation time of binaries at formation with different combinations of WDs, NSs, and BHs. The WDs are split into separate populations for helium (He), carbon/oxygen (CO), and oxygen/neon (ONe) sources. The solid lines show the formation properties of the solar metallicity population while the dashed lines show the $15\%$ solar metallicity population.} 
    \label{fig:fixed}
\end{figure*}

Furthermore, when a NS merges with another star (e.g., main-sequence star, giant, or WD) and the final product is a NS, the magnetic field and spin period of this NS are reset by drawing a new magnetic field and spin period from the same initial ranges. If a millisecond pulsar (MSP) is involved in the collision or merger, however, the newly-formed NSs are assigned different ranges of initial magnetic fields and spin periods to match those of MSPs, and the newborn NSs will remain MSPs. In this case, the magnetic fields and spin periods are randomly drawn from ranges $10^8-10^{8.8}$ G and  $3-20$ ms, respectively. In addition, we assume a lower limit of $5\times10^7$ G for the NS magnetic field \citep{Kiel2008pulsar}. No lower limit is assumed for the spin periods.

\subsection{Stellar Mergers}\label{sec:mergers}

When two stellar cores spiral in toward one another, the outcome of the subsequent merger depends upon the internal structures (i.e., density profiles) of the two objects. If one of the cores is much denser than the other (for example, if a BH/NS merges with a giant star), a common envelope-like event ensues. However, if the two stars have comparable compactness (for example, in the case of a roughly equal-mass MS--MS merger), then the merger may result in efficient mixing of the two objects. In either case, the stellar age of the merger product must be specified. Here we adopt the prescriptions of \citet{Hurley2002} to determine outcomes of stellar mergers, with one exception, outlined below.

In the case of a MS--MS merger, we assume the stellar age of the MS merger product is given by

\begin{equation}
    t_3 = f_{\rm{rejuv}} \frac{t_{\rm{MS3}}}{M_3} \Bigg( \frac{M_1 t_1}{t_{\rm{MS1}}} + \frac{M_2 t_2}{t_{\rm{MS2}}} \Bigg)
\end{equation}
where $M_1$ and $M_2$ are the masses of the two merger components, $M_3=M_1+M_2$ is the mass of the merger product \citep[assuming the stars merge without mass loss;][]{Hurley2002}, $t_{\rm{MS1}}$, $t_{\rm{MS2}}$, and $t_{\rm{MS3}}$ are the MS lifetimes of the two merger components and the merger product, respectively, and $t_1$ and $t_2$ are the stellar ages of the two merger components at the time of merger. $f_{\rm{rejuv}}$ is a factor which determines the amount of rejuvenation the merger product experiences through mixing. This factor of course depends upon the internal structure of the two stars as well as the nature of the merger (i.e., the relative velocity of the two objects at coalescence). In original \textsc{BSE}, a fixed value of 0.1 is assumed for $f_{\rm{rejuv}}$. However, in many instances this likely leads to over-rejuvenation of the merger product. Here we include $f_{\rm{rejuv}}$ as a free parameter, and adopt $f_{\rm{rejuv}}=1$ as our default value.

The outcome of stellar mergers and collisions is expected to play a critical role in dense star clusters where dynamical interactions lead to a pronounced increase in stellar mergers/collisions relative to isolated binaries \citep[e.g.,][]{HillsDay1976,Bacon1996,Lombardi2002,Fregeau2007,Leigh2011}. The details of these merger products have important implications for blue straggler stars \citep[e.g.,][]{Sandage1953,Chatterjee2013} as well as the formation of massive BHs \citep[e.g.,][]{PortegiesZwartMcMillan2002,Gurkan2004,Kremer2019}. Thus when modeling stellar and binary evolution in collisional environments like globular clusters, care must be taken when assigning the ages of stars upon collision/merger. 

We adopt the merger products from the collision matrix \cite{Hurley2002} (Table 2), but caution that the unmodified version of BSE contains a typo which causes the merger of two He-MS stars to produce a MS star. The collision matrix covers all merger types, from MS to compact remnants, and depends on the relative compactness of each component star's core. In the case of WD mergers with companion stars, the WD can mix completely with the stellar core, e.g. a merger between a He WD and a star on the first giant branch will produce a giant star with a more massive helium core. If the cores have different compactness, e.g. a merger between a CO WD and a star on the first giant branch, the product will distribute the helium core of the giant around the CO core of the white dwarf producing an early AGB star. For a detailed discussion of stellar merger products, see Section 2.7.2 of \citet{Hurley2002}.

\section{Milky Way population of compact binaries}
\label{sec:fiducialMW}

As an illustration of the capabilities of \cosmic, we simulate the Milky Way population of binaries containing combinations of WDs, NSs, and BHs with orbital periods $10\,\rm{s} < P_{\rm{orb}} < 10^{5}\,\rm{s}$.  Several population synthesis investigations using different BPS codes have predicted the population of Galactic compact binaries \citep[e.g.,][]{Hils1990,Tutukov1992,Tutukov1992b,Yungelson1995,Iben1995,Iben1995b,Iben1996,Tutukov1996,Iben1997,Nelemans2001,Nelemans2001a,Tutukov2002,Fedorova2004,Yungelson2006,Ruiter2010,Liu2014,Korol2017,Lau2019}. BPS studies have considered the impact of observed space densities \citep{Nissanke2012}, the effect of different binary evolution models on the observed population \citep{Zorotovic2010,Dominik2012,Kremer2017}, or the result of different star formation histories and Galactic spatial distributions \citep{Yu2015,Lamberts2018,Lamberts2019} have considered how Galactic compact-object binary populations are affected by different treatments of the binary evolution physics (e.g., common envelope evolution, metallicity-dependent stellar winds, etc.), different initial conditions (e.g., stellar IMF, and initial distributions of binary separation and eccentricity), and different assumptions for the Galactic SFH. As a fiducial binary evolution model, we implement several of the updated prescriptions described in \autoref{sec:upgrades} to be consistent with the models used in \cite{Kremer2019a}. In particular, we assume that compact objects are formed with the `rapid' model from \cite{Fryer2012} and that BH natal kicks are fallback modulated. We assume that ECSNe follow the \cite{Podsiadlowski2004} prescriptions and do not apply any ultra-stripped SNe prescriptions. Unless otherwise noted, we implement the defaults in the original \textsc{BSE} code release and described in Section~\ref{sec:upgrades}.

\begin{table}
	\centering
	\caption{Summary of the fixed population statistics; WD denotes the population containing He, CO, and ONe WDs. $N_{\rm{fixed}}$ is the size of the fixed population and $M_{\rm{stars}}$ is the total mass of the simulated population to produce each fixed sub-population.}
	\label{tbl:fixed}
	\begin{tabular}{lccc} 
		\hline
		 Population & $Z$ $[Z_{\odot}]$ & $\rm{N}_{\rm{fixed}}$ & $\rm{M}_{\rm{stars}}$ $[M_{\odot}]$\\
		\hline
        \hline
		He + He & 0.02 & $1.09\times10^5$ & $1.46\times10^8$\\
		CO + He & 0.02 & $1.76\times10^5$ & $1.80\times10^8$\\
		CO + CO & 0.02 & $1.24\times10^6$ & $1.13\times10^8$\\
        ONe + WD & 0.02 & $1.65\times10^5$ & $3.67\times10^8$\\
		NS + WD & 0.02 & $2.41\times10^5$ & $2.22\times10^{9}$\\
        BH + WD & 0.02 & $2.05\times10^5$ & $1.18\times10^{10}$\\
        NS + NS & 0.02 & $3.76\times10^4$ & $2.12\times10^{10}$\\
        BH + NS & 0.02 & $6.44\times10^4$ & $2.74\times10^{10}$\\
        BH + BH & 0.02 & $4.62\times10^5$ & $1.94\times10^{10}$\\
		\hline
		He + He & 0.003 & $2.29\times10^5$ & $1.31\times10^8$\\
		CO + He & 0.003 & $1.93\times10^5$ & $1.45\times10^8$\\
		CO + CO & 0.003 & $1.32\times10^6$ & $9.54\times10^7$\\
        ONe + WD & 0.003 & $2.11\times10^5$ & $3.27\times10^8$\\
		NS + WD & 0.003 & $2.76\times10^5$ & $2.37\times10^{9}$\\
        BH + WD & 0.003 & $2.12\times10^5$ & $1.13\times10^{10}$\\
        NS + NS & 0.003 & $4.35\times10^4$ & $1.55\times10^{10}$\\
        BH + NS & 0.003 & $6.76\times10^4$ & $1.28\times10^{10}$\\
        BH + BH & 0.003 & $7.58\times10^5$ & $1.33\times10^{10}$
	\end{tabular}
\end{table}

As described in \autoref{sec:cosmic}, we simulate a population for each of the Milky Way mass components: the thin and thick disks and the bulge. For the thin disk and the bulge we generate a fixed population of solar metallicity binaries from a single burst of star formation which evolves for $13.7\,\rm{Gyr}$. Similarly, we generate a fixed population of binaries with $15\%$ solar metallicity from a single burst of star formation which evolves for $13.7\,\rm{Gyr}$. We apply an upper orbital period cut of $1000\,R_{\odot}$ for systems with a WD or NS primary component since their GW merger time exceeds a Hubble time by several orders of magnitude. The number of binaries simulated in each population for both metallicities, as well as the total mass of all stars formed (including single stars) in the population are detailed in \autoref{tbl:fixed}.

\autoref{fig:fixed} shows the distributions of the masses, semilatera recta, and formation times of several combinations of WD, NS, and BH binaries at compact object formation. The solar metallicity (solid lines) populations follow similar trends when compared to the $15\%$ solar metallicity population, with the exception of the BH populations. This is primarily due to our inclusion of metallicity-dependent stellar winds, which allow for higher mass BHs. Similar to single-star evolution, the formation times of populations with a WD component are longer than those containing NS or BH components due to decreasing main sequence lifetimes with increasing progenitor mass.

\subsection{Calculation of Signal-to-Noise Ratio}
\label{sec:GW}

The characteristic strain of a GW source, as well as the signal-to-noise ratio ($S/N$) for a given GW detector, can be calculated in several different approximations, depending upon the properties of the source of interest. Specifically, the most general case of an sky and polarization averaged eccentric and chirping source can be approximated if the source is circular and/or stationary (non-chirping). In this section, we describe the computation of the characteristic strain and LISA $S/N$ in four different regimes: eccentric and chirping, circular and chirping, eccentric and stationary, and circular and stationary.

In the most general case of an eccentric chirping source, the characteristic strain at the $n$th harmonic can be written as \citep[e.g.,][]{Barack2004}:

\begin{equation}
\label{eq:hcn}
h_{c,n}^2 = \frac{1}{(\pi D_L)^2} \Bigg(\frac{2G}{c^3} \frac{\dot{E}_{n}}{\dot{f}_{n}} \Bigg).
\end{equation}
Here, $D_L$ is the luminosity distance to the source and $f_{n}$ is the source-frame GW frequency of the $n^{\rm{th}}$ harmonic given by
$f_{n} = n f_{\rm{orb}}$ where $f_{\rm{orb}}$ is the source-frame orbital frequency. $f_{n}$ is related to the observed (detector frame) GW frequency, $f_{n,\,d}$, by $f_{n} = f_{n,\,d} (1+z)$.

$\dot{E}_{n}$ is the time derivative of the energy radiated in GWs at source-frame frequency, $f_{n}$, which to lowest order is given by \citep[e.g.,][]{PetersMathews1963}:

\begin{equation}
\label{eq:Edot}
\dot{E}_{n} = \frac{32}{5} \frac{G^{7/3}}{c^5}\Big(2 \pi \,f_{\rm{orb}}\, \mathcal{M}_{c} \Big)^{10/3} \,g(n,e)
\end{equation}
where $\mathcal{M}_{c}$ is the source-frame chirp mass, which is related to detector-frame chirp mass, $\mathcal{M}_{c,\,d}$ by

\begin{figure*}
\centering
\includegraphics[width=0.9\textwidth]{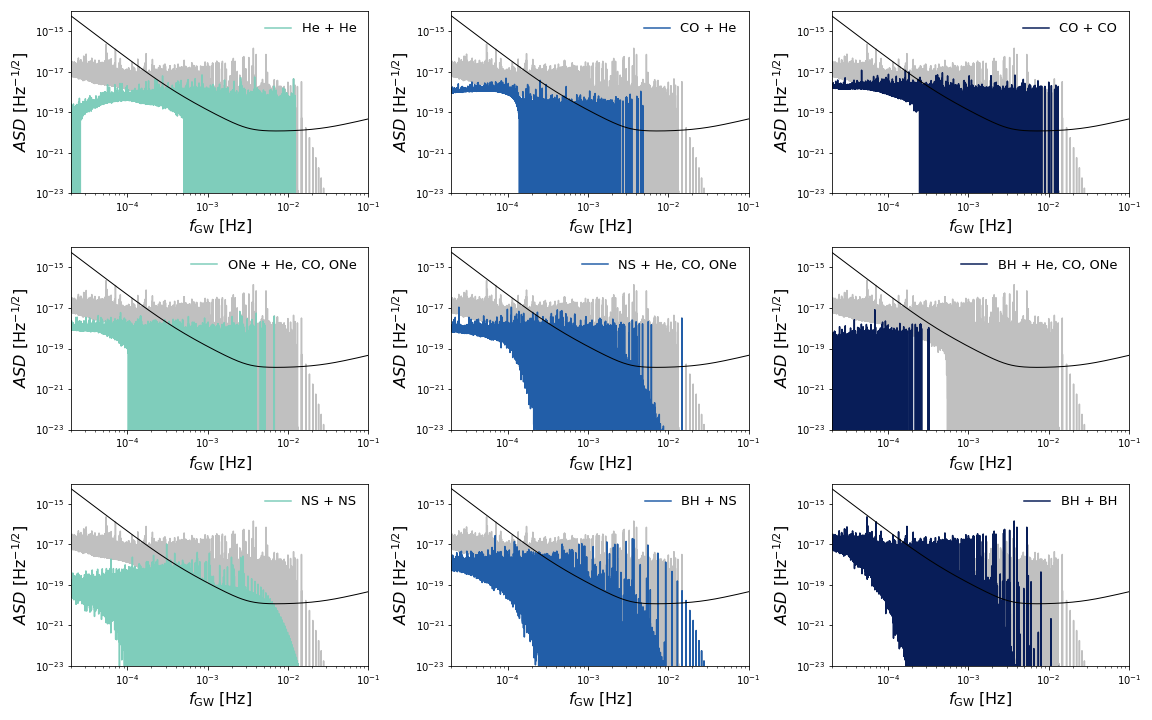}
\caption{Amplitude spectral density (ASD) as a function of frequency for each population in our Milky Way realization. The grey lines are the same in each plot and show the total ASD from the full population. The black line shows the LISA amplitude spectral density noise floor, without a Galactic foreground contribution.}
\label{fig:PSD}
\end{figure*}

\begin{equation}
\mathcal{M}_{c} =\frac{\mathcal{M}_{c,\,d}}{1+z}=\frac{(M_1M_2)^{3/5}}{(M_1+M_2)^{1/5}}\frac{1}{1+z}.
\end{equation}

$\dot{f}_{n} = n\,\dot{f}_{\rm{orb}}$ is given by:

\begin{equation}
\label{eq:fdot}
\dot{f}_{n} = n \frac{96}{10 \pi} \frac{(G\mathcal{M}_{c})^{5/3}}{c^5} \Big(2 \pi \, f_{\rm{orb}} \Big)^{11/3} \,F(e),
\end{equation}
where $F(e) = [1 + (73/24)e^2 + (37/96)e^4]/(1-e^2)^{7/2}$. Combining Equations \ref{eq:hcn}, \ref{eq:Edot}, and \ref{eq:fdot} we the obtain the characteristic strain in the detector frame:

\begin{equation}
\label{eq:strain1}
h_{c,\,n,d}^2 = \frac{2}{3\pi^{4/3}}\frac{\big(G \mathcal{M}_c \big)^{5/3}}{c^3 D_L^2} \frac{1}{f_{n,\,d}^{1/3}(1+z)^{2}}\Big( \frac{2}{n} \Big)^{2/3} \frac{g(n,e)}{F(e)}
\end{equation}

For $z \approx 0$, the distinction between the detector-frame and source-frame quantities becomes negligible. In this case, $h_{c,n,d} \approx h_{c,n}$, $f_{n,d} \approx f_n$, and $\mathcal{M}_{c,d} \approx \mathcal{M}_c$, allowing us to write Equation \ref{eq:strain1} as:

\begin{equation}
\label{eq:strain_ecc_chirp}
    h_{c,n}^2 = \frac{2}{3 \pi^{4/3}}\frac{\big(G \mathcal{M}_c \big)^{5/3}}{c^3D_L^2} \frac{1}{f_{n}^{1/3}} \Bigg( \frac{2}{n} \Bigg)^{2/3} \frac{g(n,e)}{F(e)}.
\end{equation}
Henceforth, we ignore dependence on redshift for simplicity, an appropriate approximation since $z<<1$ for the Galactic sources of interest in this analysis. For binaries with $e=0$, all GW power is emitted in the $n=2$ harmonic. Thus, the characteristic strain for circular and chirping binaries is

\begin{equation}
\label{eq:strain_hc2}
    h_{c,2}^2 = \frac{2}{3 \pi^{4/3}}\frac{\big(G \mathcal{M}_c \big)^{5/3}}{c^3D_L^2} \frac{1}{f_2^{1/3}},
\end{equation}
where we have simply taken $n=2$ in Equation \ref{eq:strain_ecc_chirp}.

For eccentric and stationary sources where $\dot{f} < f/T_{\rm{obs}}$, the dimensionless GW strain of the $n^{\rm{th}}$ harmonic is given by:

\begin{equation}
    \label{eq:h_stationary_eccentric}
    h_n^2 = \frac{128}{5} \frac{\big(G \mathcal{M}_c \big)^{10/3}}{c^8} \frac{ \big( 2\pi f_{\rm{orb}} \big)^{4/3}}{D_L^2} \frac{g(n,e)}{n^2}.
\end{equation}
Here we have divided Equation \ref{eq:strain_hc2} by two times the number of cycles, $N$, observed for a source within a given frequency bin: $N=f_n^2/\dot{f_n}$ \citep[see, e.g.,][]{Moore2015}. The strain for a stationary and circular source is obtained from Equation \ref{eq:h_stationary_eccentric} and adopting $n=2$. The amplitude spectral density ($ASD$) for a stationary source is 
\begin{equation}
    \label{eq:ASD}
    ASD_n = h_n \sqrt{T_{\rm{obs}}},
\end{equation}
\noindent where the power spectral density ($PSD$) is simply 
\begin{equation}
PSD_n = ASD_n^2.
\end{equation}

\begin{table}
	\centering
	\caption{Summary of the Milky Way population statistics; WD denotes the population containing He, CO, and ONe WDs.}
	\label{tbl:numbers}
	\begin{tabular}{lccc} 
		\hline
		Component &  Population & $\rm{N}_{\rm{total}}$ & $S/N>7$\\
		\hline
        \hline
		Thin Disk & He + He & $3.23\times10^7$ & 5053\\
		 & CO + He & $4.22\times10^7$ & 53\\
		 & CO + CO & $4.75\times10^8$ & 1684\\
         & ONe + WD & $1.94\times10^7$ & 229\\
		 & NS + WD & $4.69\times10^6$ & 240\\
         & BH + WD & $7.55\times10^5$ & 0\\
         & NS + NS & $7.67\times10^4$ & 9\\
         & BH + NS & $1.02\times10^5$ & 15\\
         & BH + BH & $1.03\times10^6$ & 62\\
		\hline
        Thick Disk & He + He & $2.53\times10^7$ & 2961\\
		 & CO + He & $1.92\times10^7$ & 10\\
		 & CO + CO & $2.00\times10^8$ & 34\\
         & ONe + WD & $9.29\times10^6$ & 12\\
		 & NS + WD & $1.68\times10^6$ & 35\\
         & BH + WD & $2.71\times10^5$ & 0\\
         & NS + NS & $4.04\times10^4$ & 1\\
         & BH + NS & $7.61\times10^4$ & 4\\
         & BH + BH & $8.22\times10^5$ & 4\\
         \hline
        Bulge & He + He & $6.65\times10^6$ & 1173\\
		 & CO + He & $8.70\times10^6$ & 2\\
		 & CO + CO & $9.79\times10^7$ & 103\\
         & ONe + WD & $4.00\times10^6$ & 10\\
		 & NS + WD & $9.66\times10^5$ & 3\\
         & BH + WD & $1.56\times10^5$ & 0\\
         & NS + NS & $1.58\times10^4$ & 0\\
         & BH + NS & $2.10\times10^4$ & 0\\
         & BH + BH & $2.12\times10^5$ & 6\\
         \hline
         \hline
         
         & Total & $2.5\times10^8$ & $1.17\times10^4$
	\end{tabular}
\end{table}

\begin{figure*}
\centering
\includegraphics[width=0.9\textwidth]{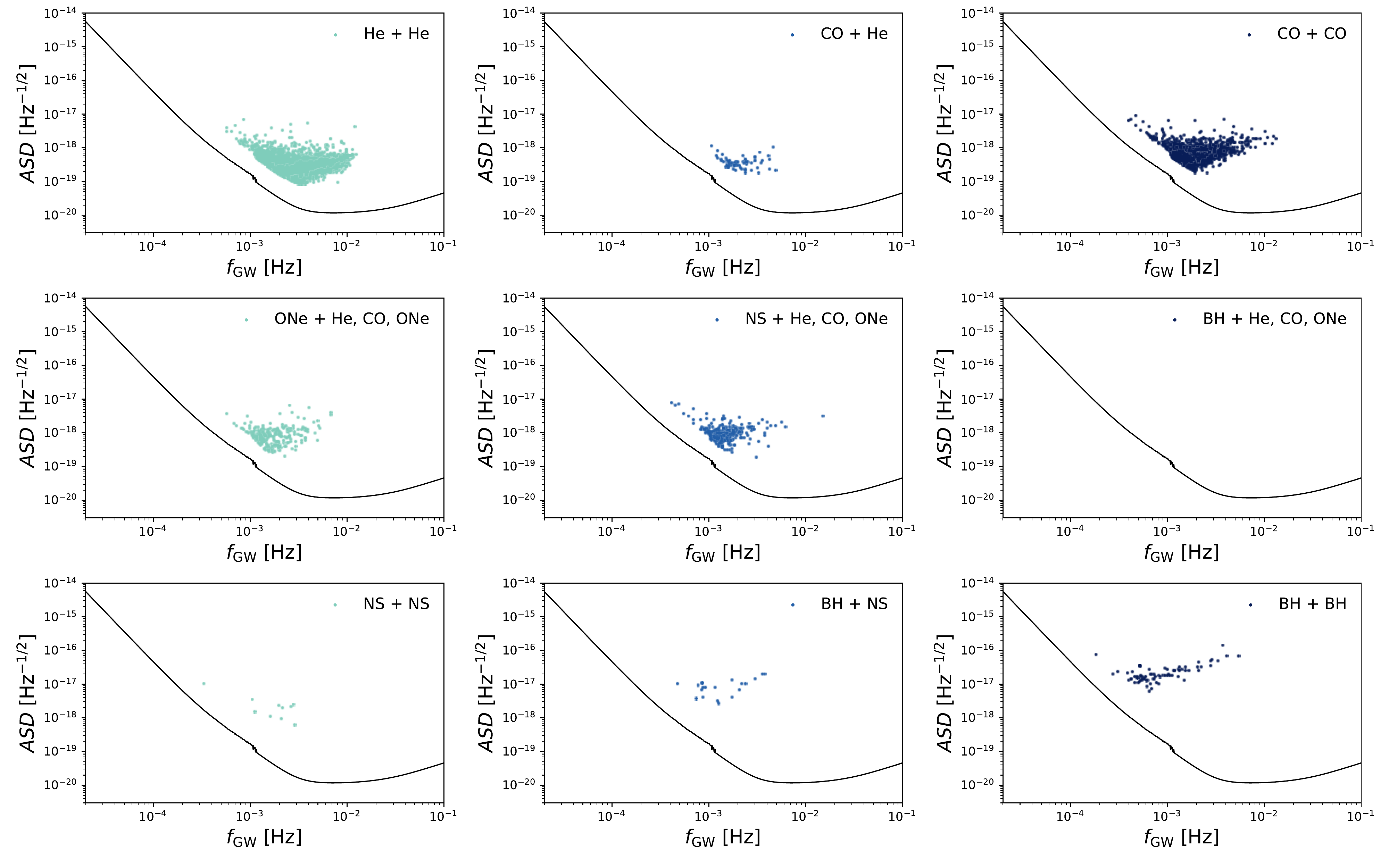}
\caption{Scatter points of the ASD vs GW frequency of the systems resolved with $S/N > 7$ for each population in our Milky Way realization. The simulated irreducible foreground and LISA sensitivity is shown in black.}
\label{fig: loud_params}
\end{figure*}

Having written the expressions for characteristic strain in the four different circular/eccentric, stationary/chirping regimes, we now go on to show how to calculate the LISA $S/N$. For chirping and eccentric sources, $S/N$ is computed following \cite{Smith2019} as
\begin{equation}
\label{eq:SNR_chirp_ecc}
\Big(\frac{S}{N} \Big)^2 = \sum_{n=1}^{\infty} \int_{f_{\rm{start}}}^{f_{\rm{end}}} \Bigg[ \frac{h_{c,\,n}(f_n)}{h_{f}(f_n)} \Bigg]^2 d \ln{f_n},
\end{equation}
where $h_{c,\,n}$ is given by Equation \ref{eq:hcn}, and $h_f$ is the characteristic LISA noise curve, which we take from \citep{Robson2019}. $f_{\rm{start}}=nf_{\rm{orb}}$ is the GW frequency emitted at the $n^{\rm{th}}$ harmonic at the start of the LISA observation and $f_{\rm{end}}$ is either the GW frequency at merger or the GW frequency of the $n^{\rm{th}}$ harmonic of the orbital frequency of the binary at the end of the LISA observation time. The characteristic noise can be expressed as
\begin{equation}
    h_f(f_n) = \sqrt{\frac{4 f_n\,P_n(f_n)}{\mathcal{R}(f_n)}},
\end{equation}
\noindent where $P_n(f_n)$ is the noise power spectral density of the detector and $\mathcal{R}(f_n)$ is the frequency-dependent signal response function. 

For chirping circular source, $h_{c,n}$ in Equation \ref{eq:SNR_chirp_ecc} is replaced by $h_{c,2}$ (Equation \ref{eq:strain_hc2}) so that the $S/N$ is given by

\begin{equation}
\label{eq:SNR_chirp_circ}
\Big(\frac{S}{N} \Big)^2 = \sum_{n=1}^{\infty} \int_{f_{\rm{start}}}^{f_{\rm{end}}} \Bigg[ \frac{h_{c,\,2}(f_n)}{h_{f}(f_n)} \Bigg]^2 d \ln{f_n}.
\end{equation}
For stationary sources, it is no longer necessary to integrate over frequency space. In the stationary and eccentric regime, we obtain:

\begin{equation}
\label{eq:SNR_stationary_ecc}
\Big(\frac{S}{N} \Big)^2 \approx \sum_{n=1}^{\infty} \Bigg[ \frac{h_{n}(f_n)}{h_{f}(f_n)} \Bigg]^2 2 f_n T_{\rm{obs}}.
\end{equation}
Finally, for stationary and circular sources:

\begin{equation}
\label{eq:SNR_stationary_circ}
\Big(\frac{S}{N} \Big)^2 \approx \Bigg[ \frac{h_{2}(f)}{h_{f}(f)} \Bigg]^2 2 f T_{\rm{obs}}.
\end{equation}
In the equations above, $T_{\rm{obs}}$ is the \LISA\ observation time which we take to be $4\,\rm{yr}$. In practice the vast majority of sources are stationary or can be approximated as stationary since their evolution over the observation time spans less than 500 \LISA\ bins. Based on a $4\,\rm{yr}$ observation time, this amounts to a frequency change of $\Delta f_{\rm{GW}} \lesssim 5\times10^{-6}\,\rm{Hz}$ and thus has a negligible effect on $S/N$. In the following section, we assume all sources are stationary.

\subsection{An example Galactic population of close WD, NS, and BH binaries}
We convolve the fixed population with the SFHs and spatial distributions described in \autoref{sec:cosmic} to produce a population of WD, NS, and BH binaries born in the Galaxy. Based on their birth time, and the age of each Galactic component, we evolve each binary up to the present and remove all systems that either merge, or fill their Roche lobes. Although accreting systems may also be resolvable GW sources, and indeed may present rich opportunities for studying various aspects of binary evolution \citep[e.g.,][]{Kremer2017,Breivik2018}, we limit this study to only detached sources for simplicity. We note that since one model is used for this example population, the uncertainty across binary evolution models is not well represented. Future studies will investigate, in detail, predicted close binary populations and the uncertainty across several models. 

\autoref{tbl:numbers} shows a summary of the number of sources per population and Milky Way component, as well as the number of systems for which the signal to noise ratio is: $S/N > 7$ from a Milky Way population realization. Our approach for selecting resolved sources is more simple than the approach used in \cite{Korol2017} and \cite{Lamberts2019}. Instead, we apply a running median with a window of $100$ frequency bins to the total root power spectral density shown in \autoref{fig:PSD} following \cite{Benacquista2006}. This produces a synthetic foreground signal similar to that shown in \cite{Korol2017}. This foreground is added to the LISA power spectral density curve of \cite{Robson2019} and used to compute the signal to noise ratio as described in \autoref{sec:GW}.

The ASD, of each population in our Milky Way realization, along with the ASD of the LISA noise floor, is plotted in \autoref{fig:PSD}. The vast majority of the signal in the ASD is due to the population of DWDs. The characteristic shape, especially the fall off near $1\,\rm{mHz}$, is due to the number of systems radiating GWs in a given frequency bin whose width is determined by LISA's observing duration as $\sim T_{\rm{obs}}^{-1}$. At low frequencies the number of sources per bin is high both because most sources preferentially have longer orbital periods and because the number of bins is low when compared with the number of bins at higher frequencies. As a result, at frequencies in excess of $1\,\rm{mHz}$, there is a higher likelihood that a source occupies a unique frequency bin. The features present in the ASD are often smoothed out to produce a foreground \citep[e.g.,][]{Littenberg2013,Korol2017, Lamberts2019}. Here, we present the unsmoothed ASD in order to illustrate contribution of each population to the signal. The GW signal at harmonics of the orbital frequency are seen in the populations containing a NS or BH due to eccentric sources. 

We note that different binary evolution models will produce different ASDs and thus different irreducible foregrounds. This may provide an avenue to distinguish different binary evolution models. However, we caution that in order to properly determine how well LISA can constrain binary evolution models, resolved sources should be compared to ASDs produced from the same fixed populations. This ensures that self-consistent comparisons are made between different models.

\autoref{fig: loud_params} shows the systems resolved with $S/N > 7$ above the irreducible foreground created from the running median of the ASD of the full population. The total number in each Galactic component is listed in \autoref{tbl:numbers}. We find that the vast majority of the population of resolved compact binaries is comprised of DWD systems. Our findings are broadly consistent with most previous work which uses population synthesis to predict LISA populations \citep[e.g.][]{Hils1990,Nelemans2001,Ruiter2010,Nissanke2012,Yu2010,Korol2017,Lamberts2018,Lau2019,Lamberts2019}. We find that the number of resolved DWDs from our simulations is in agreement within a factor of $\sim2$ \cite{Nelemans2001a,Ruiter2010,Nissanke2012,Korol2017,Lamberts2019}, however we note that we find relatively more He + He than CO + He DWDs than \cite{Lamberts2019}. Our resolved DWD population is a factor of $\sim4$ smaller DWDs than \cite{Yu2010}. We find an order of magnitude more NS + WD binaries than \cite{Nelemans2001a} resulting from updated physics for NS progenitors as described in \autoref{sec:upgrades}.

We find a lower rate of NS + NS binaries detected by \LISA\ than both \cite{Lau2019} and \cite{Andrews2019b} by a factor of $\sim3$. The assumptions for both of these studies are different from the model used here. In particular, \cite{Lau2019}, following \cite{Stevenson2017} and \cite{Vigna-Gomez2018}, assumes the initial distribution of separation is limited to $1000\,\rm{au}$ while we place an upper limit of $5000\,\rm{au}$. They also use the \cite{Fryer2012} delayed model, instead of the rapid model for compact object formation. Finally, they assume that Roche-lobe overflow from stripped stars onto NSs is stable, which produces more NS + NS binaries at relatively short orbital periods than the model used in this study. The COMPAS\_$\alpha$, detailed in \cite{Stevenson2017}, is the most similar model to the model chosen for this study and results in $\sim10$ resolved LISA sources, which is in good agreement with our findings. \cite{Andrews2019b} links the NS + NS population resolved by \LISA\ directly to the observed NS + NS merger rates derived from radio observations or \LIGOVirgo. The relative agreement of \cite{Andrews2019b, Lau2019} suggests that the model chosen for this study does not reproduce the Milky Way population of NS + NS binaries. Since this study is simply a proof of concept for the capabilities of COSMIC, we leave a careful analysis of the effects of different compact object models as a topic of future study.

Similar to \cite{Lamberts2018}, we find relatively few resolved BH + BH binaries even though there are $>10^6$ of them in our total simulated population. This is also true for BH + NS and NS + NS binaries which are produced at lower rates. The high number of BH binaries relative to NS + NS binaries is due to relationship between the mass and natal kick of the BHs and NSs at birth. In the \cite{Fryer2012} rapid model, BHs can be formed with a significant amount of fallback, and thus reduced natal kicks which are less likely to unbind the binary. However, BH binaries are mostly unresolved because they tend to occupy frequencies well below $1\,\rm{mHz}$. We further note that we use the optimistic assumption that stars which undergo a common envelope on the Hertzsprung Gap survive, leading to a larger population.

\section{Conclusion}
\label{sec:discussion}
We have presented a new, community-developed BPS code, \cosmic, which is adapted from BSE \citep{Hurley2000,Hurley2002}. We have detailed the process \cosmic\ uses to produce binary populations and shown how to scale these populations to astrophysical realizations by convolving with spatial distributions and a SFH. We have also described several updated prescriptions contained in \cosmic\ which have strong impacts for massive binary populations.

As an illustrative example, we have simulated a Milky Way realization of the thin disk, thick disk, and bulge for all combinations of stellar-remnant binaries potentially observable by \LISA. From this example population, we find $\sim10^8$ stellar remnant binaries residing in the Galaxy with $\sim10^4$ expected to be observed by \LISA\, after a $4\,\rm{yr}$ observation time, with $S/N>7$. Future studies will use \cosmic\ to self-consistently explore the resolved populations of several binary evolution models as well as Galactic star formation histories and spatial distributions. Such studies which explore the uncertainties across binary evolution models will provide important insights into compact-object populations and their progenitors, allowing comparisons to current and future GW or electromagnetic observations. 

\section*{Acknowledgements}
The authors are grateful to Vicky Kalogera who provided invaluable insights during COSMIC's development, Christopher Berry for a careful reading of the manuscript, Alejandro Vigna-Gomez, Tristan Smith, and Robert Caldwell for helpful discussions, and the anonymous referee whose comments improved the clarity of the manuscript. K.B. acknowledges Sylvia Toonen, Valeriya Korol, Astrid Lamberts, Tassos Fragos, Chris Fryer, and Stephan Justham for helpful conversations about population synthesis and support from the Jeffery L. Bishop Fellowship. J.J.A. acknowledges support by the Danish National Research Foundation (DNRF132). M.K. acknowledges support from the Illinois Space Grant Consortium. F.A.R. acknowledges support from NSF Grant AST-1716762 at Northwestern University. The majority of our analysis was performed using the computational resources of the Quest high performance computing facility at Northwestern University which is jointly supported by the Office of the Provost, the Office for Research, and Northwestern University Information Technology. 




\bibliography{references} 

\begin{thebibliography}{}
\expandafter\ifx\csname natexlab\endcsname\relax\def\natexlab#1{#1}\fi
\providecommand{\url}[1]{\href{#1}{#1}}

\bibitem[{{Abbott} {et~al.}(2018){Abbott}, {Abbott}, {Abbott}, \&
  et~al.}]{GWTC-1}
{Abbott}, B.~P., {Abbott}, R., {Abbott}, T.~D., \& et~al. 2018, arXiv e-prints,
  arXiv:1811.12907

\bibitem[{{Abbott} {et~al.}(2017){Abbott}, {Abbott}, {Abbott}, {Acernese},
  {Ackley}, {Adams}, {Adams}, {Addesso}, {Adhikari}, {Adya}, \&
  et~al.}]{GW170817}
{Abbott}, B.~P., {Abbott}, R., {Abbott}, T.~D., {et~al.} 2017, Physical Review
  Letters, 119, 161101

\bibitem[{{Andrews} {et~al.}(2019){Andrews}, {Breivik}, {Pankow}, {D'Orazio},
  \& {Safarzadeh}}]{Andrews2019b}
{Andrews}, J.~J., {Breivik}, K., {Pankow}, C., {D'Orazio}, D.~J., \&
  {Safarzadeh}, M. 2019, arXiv e-prints, arXiv:1910.13436

\bibitem[{{Andrews} {et~al.}(2015){Andrews}, {Farr}, {Kalogera}, \&
  {Willems}}]{Andrews2015}
{Andrews}, J.~J., {Farr}, W.~M., {Kalogera}, V., \& {Willems}, B. 2015, \apj,
  801, 32

\bibitem[{{Andrews} {et~al.}(2018){Andrews}, {Zezas}, \&
  {Fragos}}]{Andrews2018}
{Andrews}, J.~J., {Zezas}, A., \& {Fragos}, T. 2018, \apjs, 237, 1

\bibitem[{{Astropy Collaboration} {et~al.}(2013){Astropy Collaboration},
  {Robitaille}, {Tollerud}, {Greenfield}, {Droettboom}, {Bray}, {Aldcroft},
  {Davis}, {Ginsburg}, {Price-Whelan}, {Kerzendorf}, {Conley}, {Crighton},
  {Barbary}, {Muna}, {Ferguson}, {Grollier}, {Parikh}, {Nair}, {Unther},
  {Deil}, {Woillez}, {Conseil}, {Kramer}, {Turner}, {Singer}, {Fox}, {Weaver},
  {Zabalza}, {Edwards}, {Azalee Bostroem}, {Burke}, {Casey}, {Crawford},
  {Dencheva}, {Ely}, {Jenness}, {Labrie}, {Lim}, {Pierfederici}, {Pontzen},
  {Ptak}, {Refsdal}, {Servillat}, \& {Streicher}}]{Astropy2013}
{Astropy Collaboration}, {Robitaille}, T.~P., {Tollerud}, E.~J., {et~al.} 2013,
  \aap, 558, A33

\bibitem[{{Bacon} {et~al.}(1996){Bacon}, {Sigurdsson}, \& {Davies}}]{Bacon1996}
{Bacon}, D., {Sigurdsson}, S., \& {Davies}, M.~B. 1996, \mnras, 281, 830

\bibitem[{{Barack} \& {Cutler}(2004)}]{Barack2004}
{Barack}, L., \& {Cutler}, C. 2004, \prd, 69, 082005

\bibitem[{{Barrett} {et~al.}(2018){Barrett}, {Gaebel}, {Neijssel},
  {Vigna-G{\'o}mez}, {Stevenson}, {Berry}, {Farr}, \& {Mandel}}]{Barrett2018}
{Barrett}, J.~W., {Gaebel}, S.~M., {Neijssel}, C.~J., {et~al.} 2018, \mnras,
  477, 4685

\bibitem[{{Barrett} {et~al.}(2017){Barrett}, {Mandel}, {Neijssel}, {Stevenson},
  \& {Vigna-G{\'o}mez}}]{Barrett2017}
{Barrett}, J.~W., {Mandel}, I., {Neijssel}, C.~J., {Stevenson}, S., \&
  {Vigna-G{\'o}mez}, A. 2017, in IAU Symposium, Vol. 325, Astroinformatics, ed.
  M.~{Brescia}, S.~G. {Djorgovski}, E.~D. {Feigelson}, G.~{Longo}, \&
  S.~{Cavuoti}, 46--50

\bibitem[{{Belczynski} {et~al.}(2010){Belczynski}, {Bulik}, {Fryer}, {Ruiter},
  {Valsecchi}, {Vink}, \& {Hurley}}]{Belczynski2010}
{Belczynski}, K., {Bulik}, T., {Fryer}, C.~L., {et~al.} 2010, \apj, 714, 1217

\bibitem[{{Belczynski} {et~al.}(2002){Belczynski}, {Kalogera}, \&
  {Bulik}}]{Belczynski2002}
{Belczynski}, K., {Kalogera}, V., \& {Bulik}, T. 2002, \apj, 572, 407

\bibitem[{{Belczynski} {et~al.}(2008){Belczynski}, {Kalogera}, {Rasio}, {Taam},
  {Zezas}, {Bulik}, {Maccarone}, \& {Ivanova}}]{Belczynski2008}
{Belczynski}, K., {Kalogera}, V., {Rasio}, F.~A., {et~al.} 2008, \apjs, 174,
  223

\bibitem[{{Belczynski} {et~al.}(2016){Belczynski}, {Heger}, {Gladysz},
  {Ruiter}, {Woosley}, {Wiktorowicz}, {Chen}, {Bulik}, {O'Shaughnessy}, {Holz},
  {Fryer}, \& {Berti}}]{Belczynski2016}
{Belczynski}, K., {Heger}, A., {Gladysz}, W., {et~al.} 2016, \aap, 594, A97

\bibitem[{{Belczynski} {et~al.}(2017){Belczynski}, {Klencki}, {Fields},
  {Olejak}, {Berti}, {Meynet}, {Fryer}, {Holz}, {O'Shaughnessy}, {Brown},
  {Bulik}, {Leung}, {Nomoto}, {Madau}, {Hirschi}, {Jones}, {Mondal},
  {Chruslinska}, {Drozda}, {Gerosa}, {Doctor}, {Giersz}, {Ekstrom}, {Georgy},
  {Askar}, {Wysocki}, {Natan}, {Farr}, {Wiktorowicz}, {Miller}, {Farr}, \&
  {Lasota}}]{Belczynski2017}
{Belczynski}, K., {Klencki}, J., {Fields}, C.~E., {et~al.} 2017, arXiv
  e-prints, arXiv:1706.07053

\bibitem[{{Benacquista} \& {Holley-Bockelmann}(2006)}]{Benacquista2006}
{Benacquista}, M., \& {Holley-Bockelmann}, K. 2006, \apj, 645, 589

\bibitem[{{Bhattacharya} \& {van den Heuvel}(1991)}]{Bhattacharya1991pulsar}
{Bhattacharya}, D., \& {van den Heuvel}, E.~P.~J. 1991, \physrep, 203, 1

\bibitem[{{Bondi} \& {Hoyle}(1944)}]{BondiHoyle1944}
{Bondi}, H., \& {Hoyle}, F. 1944, \mnras, 104, 273

\bibitem[{{Breivik} {et~al.}(2018){Breivik}, {Kremer}, {Bueno}, {Larson},
  {Coughlin}, \& {Kalogera}}]{Breivik2018}
{Breivik}, K., {Kremer}, K., {Bueno}, M., {et~al.} 2018, \apjl, 854, L1

\bibitem[{Breivik {et~al.}(2019)Breivik, Coughlin, Zevin, Rodriguez, Kremer,
  Ye, Andrews, Kurkowski, Kimball, Digman, \& Tranh}]{COSMIC_3.2}
Breivik, K., Coughlin, S.~C., Zevin, M., {et~al.} 2019, COSMIC-PopSynth/COSMIC:
  COSMIC, vv3.2.0,  Zenodo, doi:10.5281/zenodo.3561144.
\newblock \url{https://doi.org/10.5281/zenodo.3561144}

\bibitem[{{Broekgaarden} {et~al.}(2019){Broekgaarden}, {Justham}, {de Mink},
  {Gair}, {Mandel}, {Stevenson}, {Barrett}, {Vigna-G{\'o}mez}, \&
  {Neijssel}}]{Broekgaarden2019}
{Broekgaarden}, F.~S., {Justham}, S., {de Mink}, S.~E., {et~al.} 2019, arXiv
  e-prints, arXiv:1905.00910

\bibitem[{{Camacho} {et~al.}(2014){Camacho}, {Torres}, {Garc{\'{\i}}a-Berro},
  {Zorotovic}, {Schreiber}, {Rebassa-Mansergas}, {Nebot G{\'o}mez-Mor{\'a}n},
  \& {G{\"a}nsicke}}]{Camacho2014}
{Camacho}, J., {Torres}, S., {Garc{\'{\i}}a-Berro}, E., {et~al.} 2014, \aap,
  566, A86

\bibitem[{{Chatterjee} {et~al.}(2013){Chatterjee}, {Rasio}, {Sills}, \&
  {Glebbeek}}]{Chatterjee2013}
{Chatterjee}, S., {Rasio}, F.~A., {Sills}, A., \& {Glebbeek}, E. 2013, \apj,
  777, 106

\bibitem[{{Chatziioannou} {et~al.}(2017){Chatziioannou}, {Clark}, {Bauswein},
  {Millhouse}, {Littenberg}, \& {Cornish}}]{Chatziioannou2017}
{Chatziioannou}, K., {Clark}, J.~A., {Bauswein}, A., {et~al.} 2017, \prd, 96,
  124035

\bibitem[{{Chen} {et~al.}(2015){Chen}, {Bressan}, {Girardi}, {Marigo}, {Kong},
  \& {Lanza}}]{Chen2015}
{Chen}, Y., {Bressan}, A., {Girardi}, L., {et~al.} 2015, \mnras, 452, 1068

\bibitem[{{Claeys} {et~al.}(2014){Claeys}, {Pols}, {Izzard}, {Vink}, \&
  {Verbunt}}]{Claeys2014}
{Claeys}, J.~S.~W., {Pols}, O.~R., {Izzard}, R.~G., {Vink}, J., \& {Verbunt},
  F.~W.~M. 2014, \aap, 563, A83

\bibitem[{{De Donder} \& {Vanbeveren}(2004)}]{DeDonder2004}
{De Donder}, E., \& {Vanbeveren}, D. 2004, \nar, 48, 861

\bibitem[{{de Kool}(1990)}]{deKool1990}
{de Kool}, M. 1990, \apj, 358, 189

\bibitem[{{de Mink} \& {Belczynski}(2015)}]{DeMink2015}
{de Mink}, S.~E., \& {Belczynski}, K. 2015, \apj, 814, 58

\bibitem[{{de Mink} {et~al.}(2007){de Mink}, {Pols}, \&
  {Hilditch}}]{deMink2007}
{de Mink}, S.~E., {Pols}, O.~R., \& {Hilditch}, R.~W. 2007, \aap, 467, 1181

\bibitem[{{Deloye} \& {Taam}(2010)}]{Deloye2010}
{Deloye}, C.~J., \& {Taam}, R.~E. 2010, \apjl, 719, L28

\bibitem[{{Dewey} \& {Cordes}(1987)}]{Dewey1987}
{Dewey}, R.~J., \& {Cordes}, J.~M. 1987, \apj, 321, 780

\bibitem[{{Dominik} {et~al.}(2012){Dominik}, {Belczynski}, {Fryer}, {Holz},
  {Berti}, {Bulik}, {Mandel}, \& {O'Shaughnessy}}]{Dominik2012}
{Dominik}, M., {Belczynski}, K., {Fryer}, C., {et~al.} 2012, \apj, 759, 52

\bibitem[{{Dominik} {et~al.}(2013){Dominik}, {Belczynski}, {Fryer}, {Holz},
  {Berti}, {Bulik}, {Mandel}, \& {O'Shaughnessy}}]{Dominik2013}
---. 2013, \apj, 779, 72

\bibitem[{{Eggenberger} {et~al.}(2008){Eggenberger}, {Meynet}, {Maeder},
  {Hirschi}, {Charbonnel}, {Talon}, \& {Ekstr{\"o}m}}]{Eggenberger2008}
{Eggenberger}, P., {Meynet}, G., {Maeder}, A., {et~al.} 2008, \apss, 316, 43

\bibitem[{{Eldridge} \& {Stanway}(2016)}]{Eldridge2016}
{Eldridge}, J.~J., \& {Stanway}, E.~R. 2016, \mnras, 462, 3302

\bibitem[{{Eldridge} {et~al.}(2017){Eldridge}, {Stanway}, {Xiao}, {McClelland
  }, {Taylor}, {Ng}, {Greis}, \& {Bray}}]{Eldridge2017}
{Eldridge}, J.~J., {Stanway}, E.~R., {Xiao}, L., {et~al.} 2017, \pasa, 34, e058

\bibitem[{{Fedorova} {et~al.}(2004){Fedorova}, {Tutukov}, \&
  {Yungelson}}]{Fedorova2004}
{Fedorova}, A.~V., {Tutukov}, A.~V., \& {Yungelson}, L.~R. 2004, Astronomy
  Letters, 30, 73

\bibitem[{{Fragos} {et~al.}(2019){Fragos}, {Andrews}, {Ramirez-Ruiz}, {Meynet},
  {Kalogera}, {Taam}, \& {Zezas}}]{Fragos2019}
{Fragos}, T., {Andrews}, J.~J., {Ramirez-Ruiz}, E., {et~al.} 2019, arXiv
  e-prints, arXiv:1907.12573

\bibitem[{{Fregeau} \& {Rasio}(2007)}]{Fregeau2007}
{Fregeau}, J.~M., \& {Rasio}, F.~A. 2007, \apj, 658, 1047

\bibitem[{Fryer {et~al.}(2012)Fryer, Belczynski, Wiktorowicz, Dominik,
  Kalogera, \& Holz}]{Fryer2012}
Fryer, C.~L., Belczynski, K., Wiktorowicz, G., {et~al.} 2012, \apj, 749, 91

\bibitem[{{Fuller} \& {Ma}(2019)}]{Fuller2019}
{Fuller}, J., \& {Ma}, L. 2019, \apjl, 881, L1

\bibitem[{{Geller} {et~al.}(2019){Geller}, {Leigh}, {Giersz}, {Kremer}, \&
  {Rasio}}]{Geller2019}
{Geller}, A.~M., {Leigh}, N. W.~C., {Giersz}, M., {Kremer}, K., \& {Rasio},
  F.~A. 2019, \apj, 872, 165

\bibitem[{{Giacobbo} \& {Mapelli}(2018)}]{Giacobbo2018b}
{Giacobbo}, N., \& {Mapelli}, M. 2018, \mnras, 480, 2011

\bibitem[{{Giacobbo} {et~al.}(2018){Giacobbo}, {Mapelli}, \&
  {Spera}}]{Giacobbo2018}
{Giacobbo}, N., {Mapelli}, M., \& {Spera}, M. 2018, \mnras, 474, 2959

\bibitem[{{Goldberg} \& {Mazeh}(1994)}]{Goldberg1994}
{Goldberg}, D., \& {Mazeh}, T. 1994, \aap, 282, 801

\bibitem[{{Gr{\"a}fener} \& {Hamann}(2008)}]{GrafenerHamann2008}
{Gr{\"a}fener}, G., \& {Hamann}, W.~R. 2008, \aap, 482, 945

\bibitem[{{G{\"u}rkan} {et~al.}(2004){G{\"u}rkan}, {Freitag}, \&
  {Rasio}}]{Gurkan2004}
{G{\"u}rkan}, M.~A., {Freitag}, M., \& {Rasio}, F.~A. 2004, \apj, 604, 632

\bibitem[{{Heggie}(1975)}]{Heggie1975}
{Heggie}, D.~C. 1975, \mnras, 173, 729

\bibitem[{{Hills} \& {Day}(1976)}]{HillsDay1976}
{Hills}, J.~G., \& {Day}, C.~A. 1976, \aplett, 17, 87

\bibitem[{{Hils} {et~al.}(1990){Hils}, {Bender}, \& {Webbink}}]{Hils1990}
{Hils}, D., {Bender}, P.~L., \& {Webbink}, R.~F. 1990, \apj, 360, 75

\bibitem[{{Hobbs} {et~al.}(2005){Hobbs}, {Lorimer}, {Lyne}, \&
  {Kramer}}]{Hobbs2005}
{Hobbs}, G., {Lorimer}, D.~R., {Lyne}, A.~G., \& {Kramer}, M. 2005, \mnras,
  360, 974

\bibitem[{{Hurley} {et~al.}(2000){Hurley}, {Pols}, \& {Tout}}]{Hurley2000}
{Hurley}, J.~R., {Pols}, O.~R., \& {Tout}, C.~A. 2000, \mnras, 315, 543

\bibitem[{{Hurley} {et~al.}(2002){Hurley}, {Tout}, \& {Pols}}]{Hurley2002}
{Hurley}, J.~R., {Tout}, C.~A., \& {Pols}, O.~R. 2002, \mnras, 329, 897

\bibitem[{{Iben} {et~al.}(1995{\natexlab{a}}){Iben}, {Tutukov}, \&
  {Yungelson}}]{Iben1995}
{Iben}, Icko, J., {Tutukov}, A.~V., \& {Yungelson}, L.~R. 1995{\natexlab{a}},
  \apjs, 100, 217

\bibitem[{{Iben} {et~al.}(1995{\natexlab{b}}){Iben}, {Tutukov}, \&
  {Yungelson}}]{Iben1995b}
---. 1995{\natexlab{b}}, \apjs, 100, 233

\bibitem[{{Iben} {et~al.}(1996){Iben}, {Tutukov}, \& {Yungelson}}]{Iben1996}
---. 1996, \apj, 456, 750

\bibitem[{{Iben} {et~al.}(1997){Iben}, {Tutukov}, \& {Yungelson}}]{Iben1997}
---. 1997, \apj, 475, 291

\bibitem[{{Ivanova} {et~al.}(2003){Ivanova}, {Belczynski}, {Kalogera}, {Rasio},
  \& {Taam}}]{Ivanova2003}
{Ivanova}, N., {Belczynski}, K., {Kalogera}, V., {Rasio}, F.~A., \& {Taam},
  R.~E. 2003, \apj, 592, 475

\bibitem[{{Ivanova} {et~al.}(2008){Ivanova}, {Heinke}, {Rasio}, {Belczynski},
  \& {Fregeau}}]{Ivanova2008}
{Ivanova}, N., {Heinke}, C.~O., {Rasio}, F.~A., {Belczynski}, K., \& {Fregeau},
  J.~M. 2008, \mnras, 386, 553

\bibitem[{{Ivanova} \& {Taam}(2004)}]{Ivanova2004}
{Ivanova}, N., \& {Taam}, R.~E. 2004, \apj, 601, 1058

\bibitem[{{Izzard} {et~al.}(2006){Izzard}, {Dray}, {Karakas}, {Lugaro}, \&
  {Tout}}]{Izzard2006}
{Izzard}, R.~G., {Dray}, L.~M., {Karakas}, A.~I., {Lugaro}, M., \& {Tout},
  C.~A. 2006, \aap, 460, 565

\bibitem[{{Izzard} {et~al.}(2009){Izzard}, {Glebbeek}, {Stancliffe}, \&
  {Pols}}]{Izzard2009}
{Izzard}, R.~G., {Glebbeek}, E., {Stancliffe}, R.~J., \& {Pols}, O.~R. 2009,
  \aap, 508, 1359

\bibitem[{{Izzard} {et~al.}(2004){Izzard}, {Tout}, {Karakas}, \&
  {Pols}}]{Izzard2004}
{Izzard}, R.~G., {Tout}, C.~A., {Karakas}, A.~I., \& {Pols}, O.~R. 2004,
  \mnras, 350, 407

\bibitem[{{Kaplan} {et~al.}(2008){Kaplan}, {Chatterjee}, {Gaensler}, \&
  {Anderson}}]{Kaplan2008}
{Kaplan}, D.~L., {Chatterjee}, S., {Gaensler}, B.~M., \& {Anderson}, J. 2008,
  \apj, 677, 1201

\bibitem[{{Kiel} {et~al.}(2008){Kiel}, {Hurley}, {Bailes}, \&
  {Murray}}]{Kiel2008pulsar}
{Kiel}, P.~D., {Hurley}, J.~R., {Bailes}, M., \& {Murray}, J.~R. 2008, \mnras,
  388, 393

\bibitem[{{Klencki} {et~al.}(2018){Klencki}, {Moe}, {Gladysz}, {Chruslinska},
  {Holz}, \& {Belczynski}}]{Klencki2018}
{Klencki}, J., {Moe}, M., {Gladysz}, W., {et~al.} 2018, \aap, 619, A77

\bibitem[{{Knuth}(2006)}]{Knuth2006}
{Knuth}, K.~H. 2006, ArXiv Physics e-prints, physics/0605197

\bibitem[{{Korol} {et~al.}(2017){Korol}, {Rossi}, {Groot}, {Nelemans},
  {Toonen}, \& {Brown}}]{Korol2017}
{Korol}, V., {Rossi}, E.~M., {Groot}, P.~J., {et~al.} 2017, \mnras, 470, 1894

\bibitem[{{Kremer} {et~al.}(2017){Kremer}, {Breivik}, {Larson}, \&
  {Kalogera}}]{Kremer2017}
{Kremer}, K., {Breivik}, K., {Larson}, S.~L., \& {Kalogera}, V. 2017, \apj,
  846, 95

\bibitem[{{Kremer} {et~al.}(2019{\natexlab{a}}){Kremer}, {Ye}, , {Rui},
  {Weatherford}, {Rodriguez}, \& et~al.}]{Kremer2019}
{Kremer}, K., {Ye}, C., , {et~al.} 2019{\natexlab{a}}

\bibitem[{{Kremer} {et~al.}(2019{\natexlab{b}}){Kremer}, {Rodriguez},
  {Amaro-Seoane}, {Breivik}, {Chatterjee}, {Katz}, {Larson}, {Rasio},
  {Samsing}, {Ye}, \& {Zevin}}]{Kremer2019a}
{Kremer}, K., {Rodriguez}, C.~L., {Amaro-Seoane}, P., {et~al.}
  2019{\natexlab{b}}, \prd, 99, 063003

\bibitem[{{Kroupa}(2001)}]{Kroupa2001}
{Kroupa}, P. 2001, \mnras, 322, 231

\bibitem[{{Kroupa} {et~al.}(1993){Kroupa}, {Tout}, \& {Gilmore}}]{Kroupa1993}
{Kroupa}, P., {Tout}, C.~A., \& {Gilmore}, G. 1993, \mnras, 262, 545

\bibitem[{{Kruckow} {et~al.}(2018){Kruckow}, {Tauris}, {Langer}, {Kramer}, \&
  {Izzard}}]{Kruckow2018}
{Kruckow}, M.~U., {Tauris}, T.~M., {Langer}, N., {Kramer}, M., \& {Izzard},
  R.~G. 2018, \mnras, 481, 1908

\bibitem[{{Lamberts} {et~al.}(2019){Lamberts}, {Blunt}, {Littenberg},
  {Garrison-Kimmel}, {Kupfer}, \& {Sanderson}}]{Lamberts2019}
{Lamberts}, A., {Blunt}, S., {Littenberg}, T.~B., {et~al.} 2019, \mnras, 2426

\bibitem[{{Lamberts} {et~al.}(2018){Lamberts}, {Garrison-Kimmel}, {Hopkins},
  {Quataert}, {Bullock}, {Faucher-Gigu{\`e}re}, {Wetzel}, {Keres}, {Drango}, \&
  {Sanderson}}]{Lamberts2018}
{Lamberts}, A., {Garrison-Kimmel}, S., {Hopkins}, P., {et~al.} 2018, ArXiv
  e-prints, arXiv:1801.03099

\bibitem[{{Lau} {et~al.}(2019){Lau}, {Mandel}, {Vigna-G{\'o}mez}, {Neijssel},
  {Stevenson}, \& {Sesana}}]{Lau2019}
{Lau}, M. Y.~M., {Mandel}, I., {Vigna-G{\'o}mez}, A., {et~al.} 2019, arXiv
  e-prints, arXiv:1910.12422

\bibitem[{{Leigh} {et~al.}(2011){Leigh}, {Sills}, \& {Knigge}}]{Leigh2011}
{Leigh}, N., {Sills}, A., \& {Knigge}, C. 2011, \mnras, 416, 1410

\bibitem[{{Lipunov} {et~al.}(1996{\natexlab{a}}){Lipunov}, {Ozernoy}, {Popov},
  {Postnov}, \& {Prokhorov}}]{Lipunov1996a}
{Lipunov}, V.~M., {Ozernoy}, L.~M., {Popov}, S.~B., {Postnov}, K.~A., \&
  {Prokhorov}, M.~E. 1996{\natexlab{a}}, \apj, 466, 234

\bibitem[{{Lipunov} \& {Postnov}(1987)}]{Lipunov1987}
{Lipunov}, V.~M., \& {Postnov}, K.~A. 1987, \sovast, 31, 228

\bibitem[{{Lipunov} {et~al.}(1996{\natexlab{b}}){Lipunov}, {Postnov}, \&
  {Prokhorov}}]{Lipunov1996b}
{Lipunov}, V.~M., {Postnov}, K.~A., \& {Prokhorov}, M.~E. 1996{\natexlab{b}},
  \aap, 310, 489

\bibitem[{{Lipunov} {et~al.}(2009){Lipunov}, {Postnov}, {Prokhorov}, \&
  {Bogomazov}}]{Lipunov2009}
{Lipunov}, V.~M., {Postnov}, K.~A., {Prokhorov}, M.~E., \& {Bogomazov}, A.~I.
  2009, Astronomy Reports, 53, 915

\bibitem[{{Littenberg} {et~al.}(2013){Littenberg}, {Larson}, {Nelemans}, \&
  {Cornish}}]{Littenberg2013}
{Littenberg}, T.~B., {Larson}, S.~L., {Nelemans}, G., \& {Cornish}, N.~J. 2013,
  \mnras, 429, 2361

\bibitem[{{Liu} \& {Zhang}(2014)}]{Liu2014}
{Liu}, J., \& {Zhang}, Y. 2014, \pasp, 126, 211

\bibitem[{{Lombardi} {et~al.}(2002){Lombardi}, {Warren}, {Rasio}, {Sills}, \&
  {Warren}}]{Lombardi2002}
{Lombardi}, James~C., J., {Warren}, J.~S., {Rasio}, F.~A., {Sills}, A., \&
  {Warren}, A.~R. 2002, \apj, 568, 939

\bibitem[{{Manchester} {et~al.}(2005){Manchester}, {Hobbs}, {Teoh}, \&
  {Hobbs}}]{Manchester2005catalogue}
{Manchester}, R.~N., {Hobbs}, G.~B., {Teoh}, A., \& {Hobbs}, M. 2005, \aj, 129,
  1993

\bibitem[{{Mapelli} {et~al.}(2009){Mapelli}, {Colpi}, \&
  {Zampieri}}]{Mapelli2009}
{Mapelli}, M., {Colpi}, M., \& {Zampieri}, L. 2009, \mnras, 395, L71

\bibitem[{{Mapelli} \& {Giacobbo}(2018)}]{Mapelli2018}
{Mapelli}, M., \& {Giacobbo}, N. 2018, \mnras, 479, 4391

\bibitem[{{Marchant} {et~al.}(2019){Marchant}, {Renzo}, {Farmer}, {Pappas},
  {Taam}, {de Mink}, \& {Kalogera}}]{Marchant2019}
{Marchant}, P., {Renzo}, M., {Farmer}, R., {et~al.} 2019, \apj, 882, 36

\bibitem[{{Mazeh} {et~al.}(1992){Mazeh}, {Goldberg}, {Duquennoy}, \&
  {Mayor}}]{Mazeh1992}
{Mazeh}, T., {Goldberg}, D., {Duquennoy}, A., \& {Mayor}, M. 1992, \apj, 401,
  265

\bibitem[{{McMillan}(2011)}]{McMillan2011}
{McMillan}, P.~J. 2011, \mnras, 414, 2446

\bibitem[{{Meynet} \& {Maeder}(2005)}]{MeynetMaeder2005}
{Meynet}, G., \& {Maeder}, A. 2005, \aap, 429, 581

\bibitem[{{Miyaji} {et~al.}(1980){Miyaji}, {Nomoto}, {Yokoi}, \&
  {Sugimoto}}]{Miyaji1980electron}
{Miyaji}, S., {Nomoto}, K., {Yokoi}, K., \& {Sugimoto}, D. 1980, \pasj, 32, 303

\bibitem[{{Moe} \& {Di Stefano}(2017)}]{Moe2017}
{Moe}, M., \& {Di Stefano}, R. 2017, \apjs, 230, 15

\bibitem[{{Moore} {et~al.}(2015){Moore}, {Cole}, \& {Berry}}]{Moore2015}
{Moore}, C.~J., {Cole}, R.~H., \& {Berry}, C.~P.~L. 2015, Classical and Quantum
  Gravity, 32, 015014

\bibitem[{{Nelemans} {et~al.}(2001){Nelemans}, {Yungelson}, {Portegies Zwart},
  \& {Verbunt}}]{Nelemans2001a}
{Nelemans}, G., {Yungelson}, L.~R., {Portegies Zwart}, S.~F., \& {Verbunt}, F.
  2001, \aap, 365, 491

\bibitem[{Nelemans {et~al.}(2001)Nelemans, Yungelson, \& Zwart}]{Nelemans2001}
Nelemans, G., Yungelson, L.~R., \& Zwart, S. F.~P. 2001, A{\&}A, 375, 890

\bibitem[{{Nelson} \& {Eggleton}(2001)}]{Nelson2001}
{Nelson}, C.~A., \& {Eggleton}, P.~P. 2001, \apj, 552, 664

\bibitem[{{Ng} \& {Romani}(2007)}]{Ng2007}
{Ng}, C.-Y., \& {Romani}, R.~W. 2007, \apj, 660, 1357

\bibitem[{{Nissanke} {et~al.}(2012){Nissanke}, {Vallisneri}, {Nelemans}, \&
  {Prince}}]{Nissanke2012}
{Nissanke}, S., {Vallisneri}, M., {Nelemans}, G., \& {Prince}, T.~A. 2012,
  \apj, 758, 131

\bibitem[{{Nomoto}(1984)}]{Nomoto1984}
{Nomoto}, K. 1984, \apj, 277, 791

\bibitem[{{Nomoto}(1987)}]{Nomoto1987}
---. 1987, \apj, 322, 206

\bibitem[{{Nomoto} \& {Kondo}(1991)}]{Nomoto1991condition}
{Nomoto}, K., \& {Kondo}, Y. 1991, \apjl, 367, L19

\bibitem[{{Paxton} {et~al.}(2011){Paxton}, {Bildsten}, {Dotter}, {Herwig},
  {Lesaffre}, \& {Timmes}}]{Paxton2011}
{Paxton}, B., {Bildsten}, L., {Dotter}, A., {et~al.} 2011, \apjs, 192, 3

\bibitem[{{Paxton} {et~al.}(2013){Paxton}, {Cantiello}, {Arras}, {Bildsten},
  {Brown}, {Dotter}, {Mankovich}, {Montgomery}, {Stello}, {Timmes}, \&
  {Townsend}}]{Paxton2013}
{Paxton}, B., {Cantiello}, M., {Arras}, P., {et~al.} 2013, \apjs, 208, 4

\bibitem[{{Paxton} {et~al.}(2015){Paxton}, {Marchant}, {Schwab}, {Bauer},
  {Bildsten}, {Cantiello}, {Dessart}, {Farmer}, {Hu}, {Langer}, {Townsend},
  {Townsley}, \& {Timmes}}]{Paxton2015}
{Paxton}, B., {Marchant}, P., {Schwab}, J., {et~al.} 2015, \apjs, 220, 15

\bibitem[{{Paxton} {et~al.}(2018){Paxton}, {Schwab}, {Bauer}, {Bildsten},
  {Blinnikov}, {Duffell}, {Farmer}, {Goldberg}, {Marchant}, {Sorokina},
  {Thoul}, {Townsend}, \& {Timmes}}]{Paxton2018}
{Paxton}, B., {Schwab}, J., {Bauer}, E.~B., {et~al.} 2018, \apjs, 234, 34

\bibitem[{{Paxton} {et~al.}(2019){Paxton}, {Smolec}, {Schwab}, {Gautschy},
  {Bildsten}, {Cantiello}, {Dotter}, {Farmer}, {Goldberg}, {Jermyn}, {Kanbur},
  {Marchant}, {Thoul}, {Townsend}, {Wolf}, {Zhang}, \& {Timmes}}]{Paxton2019}
{Paxton}, B., {Smolec}, R., {Schwab}, J., {et~al.} 2019, \apjs, 243, 10

\bibitem[{{Peters} \& {Mathews}(1963)}]{PetersMathews1963}
{Peters}, P.~C., \& {Mathews}, J. 1963, Physical Review, 131, 435

\bibitem[{{Podsiadlowski} {et~al.}(2004){Podsiadlowski}, {Langer},
  {Poelarends}, {Rappaport}, {Heger}, \& {Pfahl}}]{Podsiadlowski2004}
{Podsiadlowski}, P., {Langer}, N., {Poelarends}, A.~J.~T., {et~al.} 2004, \apj,
  612, 1044

\bibitem[{{Pols} {et~al.}(1998){Pols}, {Schr{\"o}der}, {Hurley}, {Tout}, \&
  {Eggleton}}]{Pols1998}
{Pols}, O.~R., {Schr{\"o}der}, K.-P., {Hurley}, J.~R., {Tout}, C.~A., \&
  {Eggleton}, P.~P. 1998, \mnras, 298, 525

\bibitem[{{Pols} {et~al.}(1995){Pols}, {Tout}, {Eggleton}, \& {Han}}]{Pols1995}
{Pols}, O.~R., {Tout}, C.~A., {Eggleton}, P.~P., \& {Han}, Z. 1995, \mnras,
  274, 964

\bibitem[{{Portegies Zwart} \& {McMillan}(2002)}]{PortegiesZwartMcMillan2002}
{Portegies Zwart}, S.~F., \& {McMillan}, S. L.~W. 2002, \apj, 576, 899

\bibitem[{{Portegies Zwart} \& {Verbunt}(1996)}]{Portegies1996}
{Portegies Zwart}, S.~F., \& {Verbunt}, F. 1996, \aap, 309, 179

\bibitem[{{Price-Whelan} {et~al.}(2019){Price-Whelan}, {Breivik}, {D'Orazio},
  {Hogg}, {Johnson}, {Moe}, {Morton}, \& {Tayar}}]{whitepaper}
{Price-Whelan}, A., {Breivik}, K., {D'Orazio}, D., {et~al.} 2019, \baas, 51,
  206

\bibitem[{{Rappaport} {et~al.}(1995){Rappaport}, {Podsiadlowski}, {Joss}, {Di
  Stefano}, \& {Han}}]{Rappaport1995pulsar}
{Rappaport}, S., {Podsiadlowski}, P., {Joss}, P.~C., {Di Stefano}, R., \&
  {Han}, Z. 1995, \mnras, 273, 731

\bibitem[{{Ritter} {et~al.}(1991){Ritter}, {Politano}, {Livio}, \&
  {Webbink}}]{Ritter1991}
{Ritter}, H., {Politano}, M., {Livio}, M., \& {Webbink}, R.~F. 1991, \apj, 376,
  177

\bibitem[{{Robson} {et~al.}(2019){Robson}, {Cornish}, \& {Liu}}]{Robson2019}
{Robson}, T., {Cornish}, N.~J., \& {Liu}, C. 2019, Classical and Quantum
  Gravity, 36, 105011

\bibitem[{{Rodriguez} {et~al.}(2016){Rodriguez}, {Chatterjee}, \&
  {Rasio}}]{Rodriguez2016b}
{Rodriguez}, C.~L., {Chatterjee}, S., \& {Rasio}, F.~A. 2016, PhysRevD, 93,
  084029

\bibitem[{{Ruiter} {et~al.}(2010){Ruiter}, {Belczynski}, {Benacquista},
  {Larson}, \& {Williams}}]{Ruiter2010}
{Ruiter}, A.~J., {Belczynski}, K., {Benacquista}, M., {Larson}, S.~L., \&
  {Williams}, G. 2010, \apj, 717, 1006

\bibitem[{{Saio} \& {Nomoto}(2004)}]{Saio2004off}
{Saio}, H., \& {Nomoto}, K. 2004, \apj, 615, 444

\bibitem[{{Salpeter}(1955)}]{Salpeter1955}
{Salpeter}, E.~E. 1955, ApJ, 121, 161

\bibitem[{{Sandage}(1953)}]{Sandage1953}
{Sandage}, A.~R. 1953, \aj, 58, 61

\bibitem[{{Siess} {et~al.}(2013){Siess}, {Izzard}, {Davis}, \&
  {Deschamps}}]{Siess2013}
{Siess}, L., {Izzard}, R.~G., {Davis}, P.~J., \& {Deschamps}, R. 2013, \aap,
  550, A100

\bibitem[{{Smith} \& {Caldwell}(2019)}]{Smith2019}
{Smith}, T.~L., \& {Caldwell}, R.~R. 2019, \prd, 100, 104055

\bibitem[{{Spera} \& {Mapelli}(2017)}]{Spera2017}
{Spera}, M., \& {Mapelli}, M. 2017, \mnras, 470, 4739

\bibitem[{{Spera} {et~al.}(2015){Spera}, {Mapelli}, \& {Bressan}}]{Spera2015}
{Spera}, M., {Mapelli}, M., \& {Bressan}, A. 2015, \mnras, 451, 4086

\bibitem[{{Spera} {et~al.}(2019){Spera}, {Mapelli}, {Giacobbo}, {Trani},
  {Bressan}, \& {Costa}}]{Spera2019}
{Spera}, M., {Mapelli}, M., {Giacobbo}, N., {et~al.} 2019, \mnras, 485, 889

\bibitem[{{Stanway} \& {Eldridge}(2018)}]{Stanway2018}
{Stanway}, E.~R., \& {Eldridge}, J.~J. 2018, \mnras, 479, 75

\bibitem[{{Stanway} {et~al.}(2016){Stanway}, {Eldridge}, \&
  {Becker}}]{Stanway2016}
{Stanway}, E.~R., {Eldridge}, J.~J., \& {Becker}, G.~D. 2016, \mnras, 456, 485

\bibitem[{{Stevenson} {et~al.}(2017){Stevenson}, {Berry}, \&
  {Mandel}}]{Stevenson2017}
{Stevenson}, S., {Berry}, C. P.~L., \& {Mandel}, I. 2017, \mnras, 471, 2801

\bibitem[{{Stevenson} {et~al.}(2019){Stevenson}, {Sampson}, {Powell},
  {Vigna-G{\'o}mez}, {Neijssel}, {Sz{\'e}csi}, \& {Mandel}}]{Stevenson2019}
{Stevenson}, S., {Sampson}, M., {Powell}, J., {et~al.} 2019, \apj, 882, 121

\bibitem[{{Tauris} {et~al.}(2012){Tauris}, {Langer}, \&
  {Kramer}}]{Tauris2012formation}
{Tauris}, T.~M., {Langer}, N., \& {Kramer}, M. 2012, \mnras, 425, 1601

\bibitem[{{Tauris} {et~al.}(2013){Tauris}, {Langer}, {Moriya}, {Podsiadlowski},
  {Yoon}, \& {Blinnikov}}]{Tauris2013}
{Tauris}, T.~M., {Langer}, N., {Moriya}, T.~J., {et~al.} 2013, \apjl, 778, L23

\bibitem[{{Tauris} {et~al.}(2015){Tauris}, {Langer}, \&
  {Podsiadlowski}}]{Tauris2015}
{Tauris}, T.~M., {Langer}, N., \& {Podsiadlowski}, P. 2015, \mnras, 451, 2123

\bibitem[{{Taylor} \& {Gerosa}(2018)}]{Taylor2018}
{Taylor}, S.~R., \& {Gerosa}, D. 2018, \prd, 98, 083017

\bibitem[{{Toonen} {et~al.}(2014){Toonen}, {Claeys}, {Mennekens}, \&
  {Ruiter}}]{Toonen2014}
{Toonen}, S., {Claeys}, J.~S.~W., {Mennekens}, N., \& {Ruiter}, A.~J. 2014,
  \aap, 562, A14

\bibitem[{{Toonen} \& {Nelemans}(2013)}]{Toonen2013}
{Toonen}, S., \& {Nelemans}, G. 2013, \aap, 557, A87

\bibitem[{{Toonen} {et~al.}(2012){Toonen}, {Nelemans}, \& {Portegies
  Zwart}}]{Toonen2012}
{Toonen}, S., {Nelemans}, G., \& {Portegies Zwart}, S. 2012, \aap, 546, A70

\bibitem[{{Tutukov} \& {Yungelson}(1996)}]{Tutukov1996}
{Tutukov}, A., \& {Yungelson}, L. 1996, \mnras, 280, 1035

\bibitem[{{Tutukov} \& {Yungelson}(1992)}]{Tutukov1992}
{Tutukov}, A.~V., \& {Yungelson}, L.~R. 1992, \sovast, 36, 266

\bibitem[{{Tutukov} \& {Yungelson}(2002)}]{Tutukov2002}
---. 2002, Astronomy Reports, 46, 667

\bibitem[{{Tutukov} {et~al.}(1992){Tutukov}, {Yungelson}, \&
  {Iben}}]{Tutukov1992b}
{Tutukov}, A.~V., {Yungelson}, L.~R., \& {Iben}, Icko, J. 1992, \apj, 386, 197

\bibitem[{{van Haaften} {et~al.}(2013){van Haaften}, {Nelemans}, {Voss},
  {Toonen}, {Portegies Zwart}, {Yungelson}, \& {van der
  Sluys}}]{vanHaaften2013}
{van Haaften}, L.~M., {Nelemans}, G., {Voss}, R., {et~al.} 2013, \aap, 552, A69

\bibitem[{{Vigna-G{\'o}mez} {et~al.}(2018){Vigna-G{\'o}mez}, {Neijssel},
  {Stevenson}, {Barrett}, {Belczynski}, {Justham}, {de Mink}, {M{\"u}ller},
  {Podsiadlowski}, {Renzo}, {Sz{\'e}csi}, \& {Mandel}}]{Vigna-Gomez2018}
{Vigna-G{\'o}mez}, A., {Neijssel}, C.~J., {Stevenson}, S., {et~al.} 2018,
  \mnras, 481, 4009

\bibitem[{Vink \& de~Koter(2005)}]{Vink2005}
Vink, J.~S., \& de~Koter, A. 2005, \aap, 442, 587

\bibitem[{Vink {et~al.}(2001)Vink, de~Koter, \& Lamers}]{Vink2001}
Vink, J.~S., de~Koter, A., \& Lamers, H. J. G. L.~M. 2001, \rmxaa, 369, 574

\bibitem[{{Vink} {et~al.}(2011){Vink}, {Muijres}, {Anthonisse}, {de Koter},
  {Gr{\"a}fener}, \& {Langer}}]{Vink2011}
{Vink}, J.~S., {Muijres}, L.~E., {Anthonisse}, B., {et~al.} 2011, \aap, 531,
  A132

\bibitem[{{Wang} {et~al.}(2006){Wang}, {Lai}, \& {Han}}]{Wang2006}
{Wang}, C., {Lai}, D., \& {Han}, J.~L. 2006, \apj, 639, 1007

\bibitem[{{Webbink}(1985)}]{Webbink1985}
{Webbink}, R.~F. 1985, {Stellar evolution and binaries}, ed. J.~E. {Pringle} \&
  R.~A. {Wade}, 39

\bibitem[{{Whyte} \& {Eggleton}(1985)}]{Whyte1985}
{Whyte}, C.~A., \& {Eggleton}, P.~P. 1985, \mnras, 214, 357

\bibitem[{{Woosley}(2017)}]{Woosley2017}
{Woosley}, S.~E. 2017, \apj, 836, 244

\bibitem[{{Woosley}(2019)}]{Woosley2019}
---. 2019, \apj, 878, 49

\bibitem[{{Woosley} \& {Heger}(2015)}]{Woosley2015}
{Woosley}, S.~E., \& {Heger}, A. 2015, in Astrophysics and Space Science
  Library, Vol. 412, Very Massive Stars in the Local Universe, ed. J.~S.
  {Vink}, 199

\bibitem[{{Ye} {et~al.}(2019){Ye}, {Kremer}, {Chatterjee}, {Rodriguez}, \&
  {Rasio}}]{ye2019msp}
{Ye}, C.~S., {Kremer}, K., {Chatterjee}, S., {Rodriguez}, C.~L., \& {Rasio},
  F.~A. 2019, \apj, 877, 122

\bibitem[{{Yu} \& {Jeffery}(2010)}]{Yu2010}
{Yu}, S., \& {Jeffery}, C.~S. 2010, \aap, 521, A85

\bibitem[{{Yu} \& {Jeffery}(2015)}]{Yu2015}
---. 2015, \mnras, 448, 1078

\bibitem[{{Yungelson} {et~al.}(1995){Yungelson}, {Livio}, {Tutukov}, \&
  {Kenyon}}]{Yungelson1995}
{Yungelson}, L., {Livio}, M., {Tutukov}, A., \& {Kenyon}, S.~J. 1995, \apj,
  447, 656

\bibitem[{{Yungelson} {et~al.}(2006){Yungelson}, {Lasota}, {Nelemans}, {Dubus},
  {van den Heuvel}, {Dewi}, \& {Portegies Zwart}}]{Yungelson2006}
{Yungelson}, L.~R., {Lasota}, J.~P., {Nelemans}, G., {et~al.} 2006, \aap, 454,
  559

\bibitem[{{Zevin} {et~al.}(2019){Zevin}, {Kremer}, {Siegel}, {Coughlin},
  {Tsang}, {Berry}, \& {Kalogera}}]{Zevin2019}
{Zevin}, M., {Kremer}, K., {Siegel}, D.~M., {et~al.} 2019, arXiv e-prints,
  arXiv:1906.11299

\bibitem[{{Zorotovic} {et~al.}(2010){Zorotovic}, {Schreiber}, {G{\"a}nsicke},
  \& {Nebot G{\'o}mez-Mor{\'a}n}}]{Zorotovic2010}
{Zorotovic}, M., {Schreiber}, M.~R., {G{\"a}nsicke}, B.~T., \& {Nebot
  G{\'o}mez-Mor{\'a}n}, A. 2010, \aap, 520, A86

\end{thebibliography}




%
%
%

\end{document}